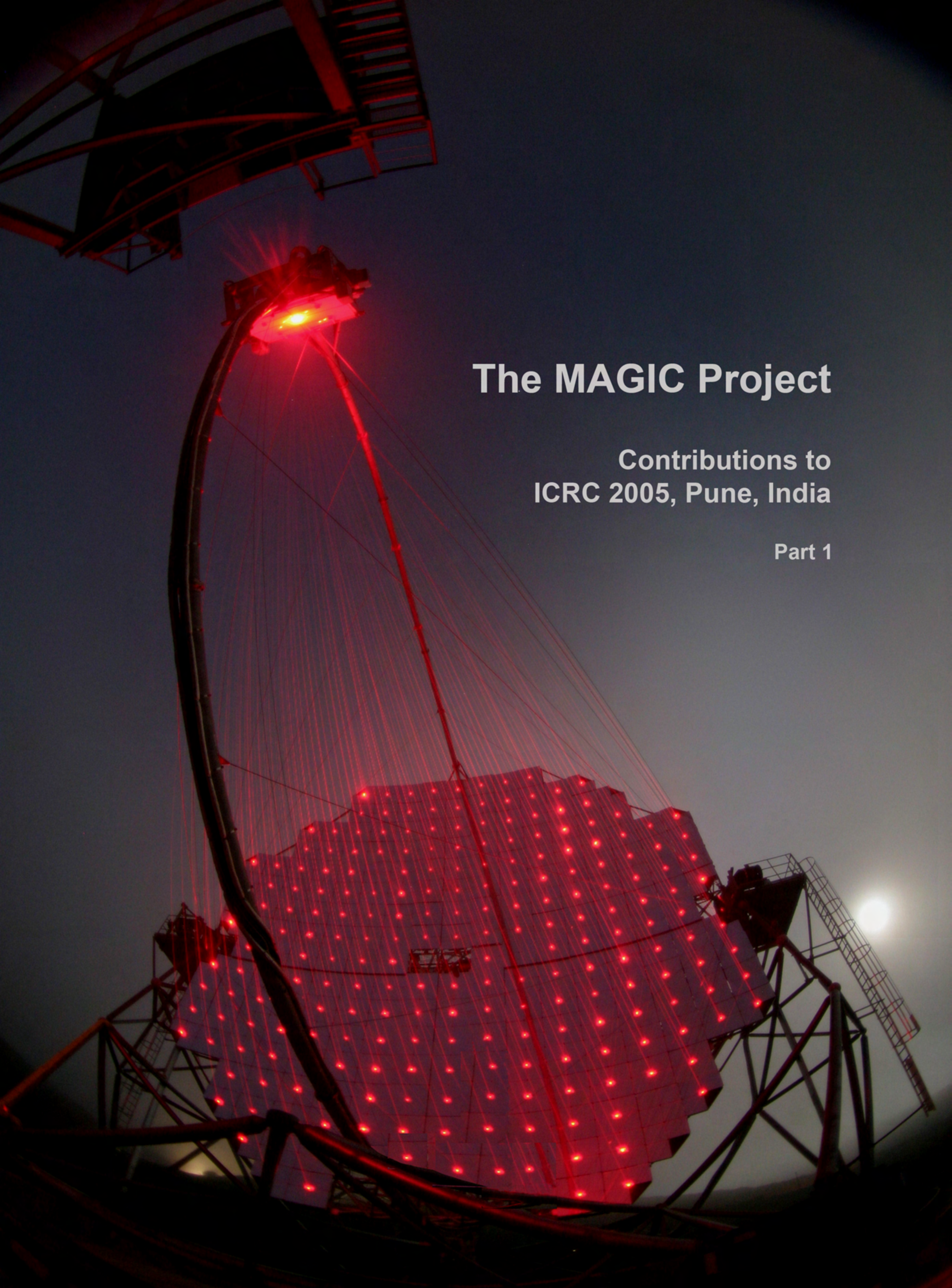

# The MAGIC Project

## Contributions to ICRC 2005, Pune, India

Part 1

## Contents

# MAGIC: List of collaboration members

J. Albert i Fort [a], E. Aliu [b], H. Anderhub [g], P. Antoranz [k], A. Armada [b], M. Asensio [k], C. Baixeras [c], J.A. Barrio [k], H. Bartko [d], D. Bastieri [e], W. Bednarek [f], K. Berger [a], C. Bigongiari [e], A. Biland [g], E. Bisesi [h], O. Blanch [b], R.K. Bock [d], T. Bretz [a], I. Britvitch [g], M. Camara [k], A. Chilingarian [i], S. Ciprini [j], J.A. Coarasa [d], S. Commichau [g], J.L. Contreras [k], J. Cortina [b], V. Danielyan [i], F. Dazzi [e], A. De Angelis [h], B. De Lotto [h], E. Domingo [b], D. Dorner [a], M. Doro [b], O. Epler [l], M. Errando [b], D. Ferenc [m], E. Fernandez [b], R. Firpo [b], J. Flix [b], M.V. Fonseca [k], L. Font [c], N. Galante [n], M. Garczarczyk [d], M. Gaug [b], J. Gebauer [d], R. Giannitrapani [h], M. Giller [f], F. Goebel [d], D. Hakobyan [i], M. Hayashida [d], T. Hengstebeck [l], D. Höhne [a], J. Hose [d], P. Jacon [f], O.C. de Jager [o], O. Kalekin [l], D. Kranich [m], A. Laille [m], T. Lenisa [h], P. Liebing [d], E. Lindfors [j], F. Longo [h], M. Lopez [k], J. Lopez [b], E. Lorenz [d, g], F. Lucarelli [k], P. Majumdar [d], G. Maneva [q], K. Mannheim [a], M. Mariotti [e], M. Martinez [b], K. Mase [d], D. Mazin [d], C. Merck [d], M. Merck [a], M. Meucci [n], M. Meyer [a], J.M. Miranda [k], R. Mirzoyan [d], S. Mizobuchi [d], A. Moralejo [e], E. Ona-Wilhelmi [b], R. Orduna [c], N. Otte [d], I.Oya [k], D. Paneque [d], R. Paoletti [n], M. Pasanen [j], D. Pascoli [e], F. Pauss [g], N. Pavel [l], R. Pegna [n], L. Peruzzo [e], A. Piccioli [n], M. Pin [h], E. Prandini [e], R. de los Reyes [k], J. Rico [b], W. Rhode [p], B. Riegel [a], M. Rissi [g], A. Robert [c], G. Rossato [e], S. Rügamer [a], A. Saggion [e], A. Sanchez [e], P. Sartori [e], V. Scalzotto [e], R. Schmitt [a], T. Schweizer [l], M. Shayduk [l], K. Shinozaki [d], N. Sidro [b], A. Sillanpää [j], D. Sobczynska [f], A. Stamerra [n], L. Stark [g], L. Takalo [j], P. Temnikov [q], D. Tescaro [e], M. Teshima [d], N. Tonello [d], A. Torres [c], N. Turini [n], H. Vankov [q], V. Vitale [d], S. Volkov [l], R. Wagner [d], T. Wibig [f], W. Wittek [d], J. Zapatero [c]

[a] Universität Würzburg, Germany
[b] Institut de Fisica d'Altes Energies, Barcelona, Spain
[c] Universitat Autonoma de Barcelona, Spain
[d] Max-Planck-Institut für Physik, München, Germany
[e] Dipartimento di Fisica, Università di Padova, and INFN Padova, Italy
[f] Division of Experimental Physics, University of Lodz, Poland
[g] Institute for Particle Physics, ETH Zürich, Switzerland
[h] Dipartimento di Fisica, Università di Udine, and INFN Trieste, Italy
[i] Yerevan Physics Institute, Cosmic Ray Division, Yerevan, Armenia
[j] Tuorla Observatory, Pikkiö, Finland
[k] Universidad Complutense, Madrid, Spain
[l] Institut für Physik, Humboldt-Universität Berlin, Germany
[m] University of California, Davis, USA
[n] Dipartimento di Fisica, Università di Siena, and INFN Pisa, Italy

[o] Space Research Unit, Potchefstroom University, South Africa
[p] Fachbereich Physik, Universität Dortmund, Germany
[q] Institute for Nuclear Research and Nuclear Energy, Sofia, Bulgaria

# Part 1: Observations

# Observations of VHE Gamma Radiation from HESS J 1813-178 with the MAGIC Telescope

H. Bartko[a], T. Bretz[b], W. Bednarek[c], J. Cortina[c], F. Goebel[a], M. Gaug[d], M. Hayashida[a], M. Mariotti[e], D. Mazin[a], R. Mirzoyan[a], S. Mizobuchi[a], A. Moralejo[e], E. de Oña[d], N. Otte[a], J. Rico[d], T. Schweizer[f], M. Teshima[a], D. Torres[d], W. Wittek[a] for the MAGIC collaboration

*(a) Max-Planck-Institute for Physics, Munich Germany*
*(b) University of Würzburg, Germany*
*(c) University of Lodz, Poland*
*(d) Institut de Fisica d Altes Energies, Edifici Cn Universitat Autonoma de Barcelona, Bellaterra, Spain*
*(e) University and INFN Padova, Italy*
*(f) Humboldt University Berlin, Germany.*

Presenter: H. Bartko (hbartko@mppmu.mpg.de), ger-bartko-H-abs4-xx-poster

Recently, gamma radiation above a few hundred GeV has been detected from eight new sources in the Galactic Plane by the HESS collaboration [1]. The source HESS J 1813-178 (HESS1813) has caused particular interest as its nature was unknown. Subsequent radio observations imply an association with the SNR G12.82-0.02 [2]. The source has also been detected by the INTEGRAL and ASCA satellites [3, 4]. At La Palma this source culminates at a zenith angle (ZA) of about 47°. MAGIC has observed the source up to 52° ZA, for 20 hours in June and July 2005. The large zenith angle observations lead to a good sensitivity for energies in the TeV range. The observation confirms the VHE gamma emission from HESS J 1813-178. A preliminary flux determination yields a flux above 300 GeV of about 6% of the Crab flux with a rather hard spectrum.

## 1. Introduction

In the galactic plane scan performed in 2004 by HESS [1], eight new sources have been detected. One of the newly detected gamma-ray sources is HESS J 1813-178 (HESS1813). Initially it was assumed to be a “dark particle accelerator” as no counter parts at other wavelengths had been identified.

Since the original discovery, HESS1813 has been associated with the SNR G12.82-0.02 [3, 4, 2]. Figure 1 shows a VLA 20 cm image of the region around G12.82-0.02 [2]. The location of HESS1813 is indicated as the smaller circle. The chance probability for spatial coincidences with a SNR in this region is non-negligible (6%) [1].

| | |
|---|---|
| (RA, dec), epoch J2000.0 | $(18^h 13^m 23^s, -17°56')$ |
| heliocentric distance | $\geq 4$ kpc (1 deg = 70 pc) |
| shell diameter | 0.03 deg |
| observation zenith angle at La Palma | $47° \leq$ ZA $\leq 52°$ |

**Table 1.** Observational parameters of SNR G12.82-0.02.

HESS J1813-178 has been found to be nearly point like. A brightness distribution $\rho = exp(-r^2/\sigma^2)$ with a size $\sigma = 3'$ has been reported [1]. The source lies at 10 arc min distance from the center of the radio source W 33. W 33 has an extension of at least 15 arc min, with a compact radio core (G12.8-0.2) that is 1 arc min across [5]. This patch in the sky is highly obscured and has indications of being a recent star formation region [6].

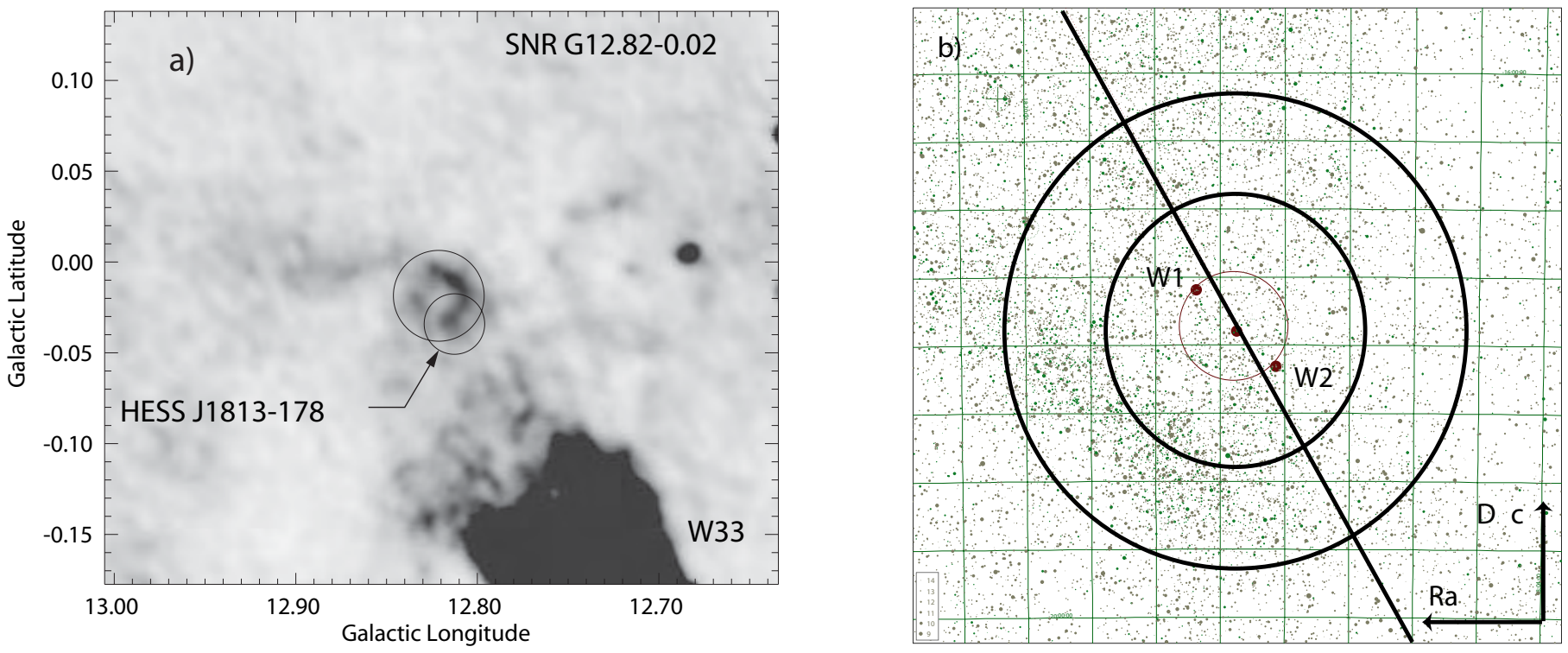

**Figure 1.** *(a) VLA 20 cm image of SNR G12.82-0.02 (indicated by the larger circle). The bright nearby HII region W33 can be seen at the lower right [2]. (b) The star field around HESS1813. Stars up to a magnitude of 14 are shown. The two big circles correspond to distances of 1° and 1.75° from HESS1813, respectively. The wobble positions W1 and W2 are given by the full circles. The $x$ axis is pointing to the direction of decreasing RA, the $y$ axis into the direction of increasing declination. The grid spacing in declination is 0.5 degree.*

## 2. Observations

The Major Atmospheric Imaging Cherenkov telescope (MAGIC [7, 8]) is currently the largest single dish Imaging Air Cherenkov Telescope (IACT). Located on the Canary Island La Palma ($28.8\,°$N, $17.8°$W, 2200m a.s.l), the telescope has a 17m diameter tessellated parabolic mirror, supported by a light weight carbon fiber frame. It is equipped with a high efficiency 576-pixel $3.5\,°$ FOV photomultiplier camera. The analogue signals are transported via optical fibers to the trigger electronics and are read out by a 300 MSamples/s FADC system.

In La Palma HESS1813 culminates at about $47\,°$ ZA. The large ZA implies a higher energy threshold for MAGIC, of about 300 GeV, but it also provides a larger effective collection area [9]. This makes sensitive measurements in the multi-TeV energy regime possible.

The sky region around HESS1813 has a relatively high and non-uniform level of light of the night sky, see figure 1b). Within a distance of $1\,°$ from HESS1813 there are no stars brighter than mag = 8. In the region south west of the source the star field is brighter. This together with the large ZA requires either observations in the false-source tracking (wobble) mode [10] or to take dedicated OFF data. The sky directions (W1, W2) to be tracked in the wobble mode are chosen such that in the camera the sky field relative to the source position (HESS1813) is similar for both wobble positions, see figure 1b). The source direction is $0.4\,°$ offset from the camera center. During wobble mode data taking, 50% of the data is taken at W1 and 50% at W2, switching (wobbling) between the 2 directions every 30 minutes.

## 3. Data Analysis

HESS1813 was observed for a total of about 20 hours in June-July 2005 in the so called wobble mode (ZA $\leq 52°$). In total, about 15 million triggers have been recorded. Image cleaning tail cuts have been applied: Pixels are only considered to be part of the image if their reconstructed charge signal is larger than 10 photo electrons (core pixels) or larger than 5 photo electrons (boundary pixel) and they are neighbors of a core pixel. After filter cuts about 10 million events remained. These data were processed for the $\gamma$/hadron separation.

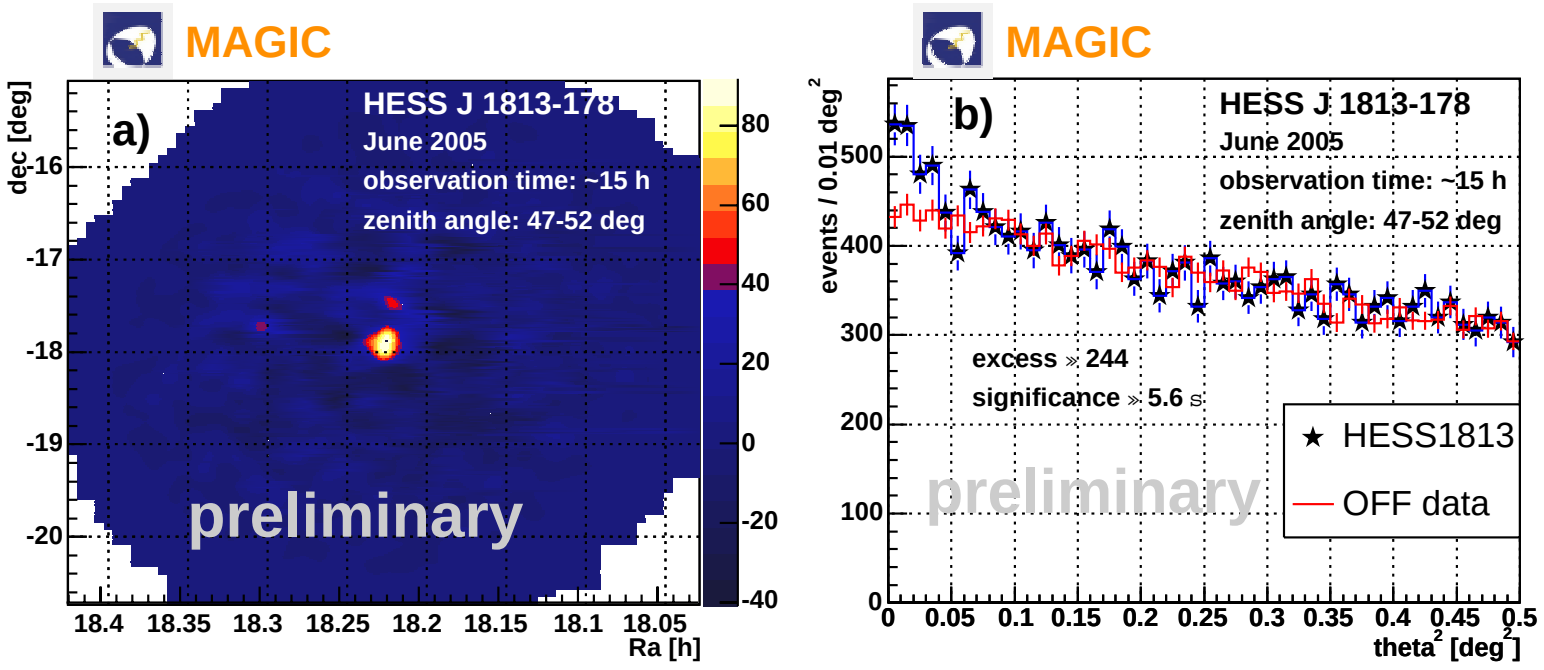

**Figure 2.** *(a) Sky map of candidate gamma-ray excess events in the directions of HESS1813. (b) Distributions of* $\theta^2$ *values for the source and anti-source, see text, for SIZE* $\geq 300$ *photo electrons (corresponding to about 300 GeV) and a hadronness*$< 0.15$.

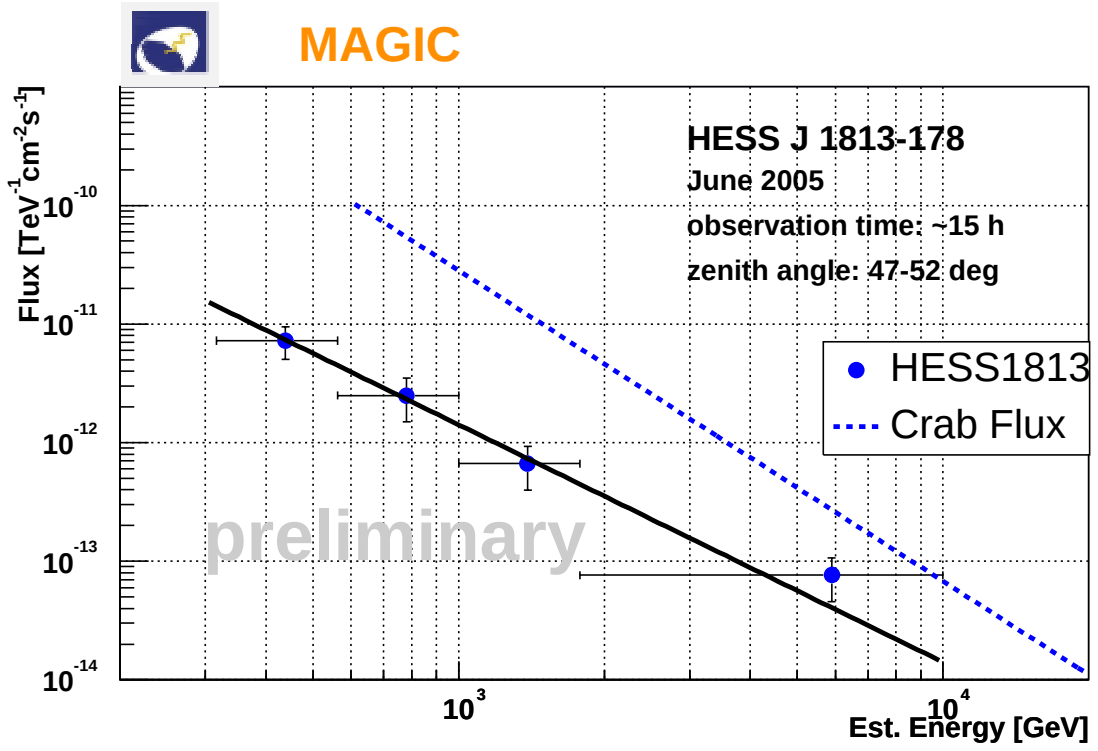

**Figure 3.** *Reconstructed VHE gamma-ray spectrum of HESS1813. The spectral index is* $-2.0 \pm 0.8$ *and the integral flux above 315 GeV is about 6% of the Crab nebula (statistical errors only). The upper curve shows the spectrum of the Crab nebula as measured by MAGIC [14].*

We performed this analysis similar to the one described in [11]. We used the Random Forest method [12] for the gamma hadron separation and the energy estimation. Every event is assigned an energy value and a parameter called hadronness $\in [0; 1]$ which is a measure for the probability to be a background event.

To train the Random Forest algorithm, high ZA (50° ZA) Monte Carlo (MC) gamma showers were generated with energies between 200 GeV and 30 TeV. The spectral index of the generated differential spectrum is $-2.6$, conforming with the energy spectrum of the Crab nebula. The MC sample was divided into two sub-samples. The algorithm was trained with one sub-sample of the MC gammas and a sub-sample of the wobble data which contains mainly background events with a negligible contamination with gammas. Subsequently, the algorithm was tested using the remaining samples of MC and wobble data. As training parameters we used the image parameters SIZE, WIDTH, LENGTH, CONC, and M3Long [13].

In order to develop and verify the analysis at high zenith angles, Crab data in the interesting ZA range around 50° have been taken in January 2005. From that sample, we determined the energy spectrum and found it to be consistent with existing other measurements, see figure 3, upper curve.

For each event the sky position is determined by using the so-called DISP-method [10, 16]. Figure 2a) shows the sky map of gamma-ray candidate excess events from the direction of HESS1813 with a lower SIZE cut of 300 photo electrons and a hadronness $< 0.15$. The size cut corresponds to an energy threshold of about 300 GeV. Figure 2b) shows the distribution of the squared angular distance, $\theta^2$, between the reconstructed shower direction and the nominal object position. The observed excess in the direction of HESS1813 is compatible within errors with the measurement of HESS [1]. It is also compatible with a point source like emission.

Figure 3 shows the reconstructed VHE gamma-ray spectrum of HESS1813. No unfolding with the energy resolution of the detector has been applied yet. The spectral index ($\mathrm{d}N/(\mathrm{d}E\mathrm{d}A\mathrm{d}t) \sim E^\Gamma$) was found to be $\Gamma = -2.0 \pm 0.8$ and the integral flux above 315 GeV corresponds to about 6% of the Crab nebula (statistical errors only). The results are still preliminary; the final analysis will be presented elsewhere.

## 4. Conclusions

The detection of HESS J 1813-178 using the MAGIC Telescope confirms a new VHE gamma-ray source in the Galactic Plane. A reasonably large data set was collected from observations at large zenith angles to infer the spectrum of this source up to energies of about 6 TeV. Between 300 GeV and 6 TeV the differential energy spectrum can be fitted with a power law of slope $\Gamma = -2.0 \pm 0.8$. The data can be used to cross-calibrate the HESS and MAGIC IACTs, and show satisfactory agreement.

Observations in the radio, X-ray and gamma-ray band imply a connection between HESS J 1813-178 and the SNR G12.82-0.02 [2, 3, 4]. Generally, hard gamma-ray spectra are expected from SNRs due to Fermi acceleration of cosmic rays [15]. The hard spectrum determined for HESS J 1813-178 may be a further hint for its association with the SNR G12.82-0.02.

Surveying the inventory of galactic VHE gamma ray sources puts models for the origin of cosmic rays into a new perspective. In particular, modeling the source distribution for propagation models that predict the Galactic diffuse emission will provide added value to these observations in the GLAST era.

## Acknowledgements

The project MAGIC is supported by MPG and BMBF (Germany), INFN (Italy), CICYT and IAC (Spain).

# Search for Gamma Rays from the Galactic Center with the MAGIC Telescope

H. Bartko[a], A. Biland[b], E. Bisesi[c], S. Commichau[b], P. Flix[d], E. Lorenz[a,b], M. Mariotti[e], R. Mirzoyan[a], V. Scalzotto[e], S. Stark[b], W. Wittek[a] for the MAGIC collaboration
*(a) Max-Planck-Institute for Physics, Munich Germany*
*(b) ETH Zurich, Switzerland*
*(c) University of Udine and INFN Trieste, Italy*
*(d) Institut de Fisica d Altes Energies, Edifici Cn Universitat Autonoma de Barcelona, Bellaterra, Spain*
*(e) University and INFN Padova, Italy*
Presenter: H. Bartko (hbartko@mppmu.mpg.de), ger-bartko-H-abs2-og22-oral

The Galactic Center (GC) is a very interesting region for gamma ray astronomy. Various possibilities for the production of very high energy (VHE) gamma rays near the GC are discussed in the literature, like accretion flow onto the central black hole, supernova shocks in Sgr A East, proton acceleration near the event horizon of the black hole, or WIMP dark matter annihilation. At the Canary Island La Palma, the GC culminates at about 58 deg zenith angle (ZA). Between May and August it can be observed with the MAGIC telescope at up to 60 deg ZA. The large zenith angle leads to a good sensitivity for energies in the TeV range. The observation and analysis strategies are outlined and the status of the ongoing analysis is presented.

## 1. Introduction

The Galactic Center (GC) region contains many unusual objects which may be responsible for high-energy processes generating gamma rays [1, 2, 3]. The GC is rich in massive stellar clusters with up to 100 OB stars [4], immersed in a dense gas. There are young supernova remnants e.g. G0.570-0.018 or Sgr A East, and nonthermal radio arcs. The dynamical center of the Milky Way is associated with the compact radio source Sgr A$^*$, which is believed to be a massive black hole [4]. Within a radius of 300 pc around the Galactic Center there is a mass of about $3 \cdot 10^7 M_{\odot}$. Some data about the GC are summarized in Table 1.

| | |
|---|---|
| (RA, dec), epoch J2000.0 | $(17^h 45^m 36^s, -28°56')$ |
| heliocentric distance | $8 \pm 0.4$ kpc (1 deg = 140 pc) |
| mass of the black hole | $2 \pm 0.5 \cdot 10^6 M_{\odot}$ |
| min. zenith angle at La Palma | 58 deg |

**Table 1.** Properties of the Galactic Center.

EGRET has detected a strong source in the direction of the GC, 3 EG J1745-2852 [5], which has a broken power law spectrum extending up to at least 10 GeV, with a spectral index of 1.3 below the break at a few GeV. Assuming a distance of 8.5 kpc, the gamma ray luminosity of this source is very large, $2.2 \cdot 10^{37} \mathrm{erg/s}$, which is equivalent to about 10 times the gamma flux from the Crab nebula. An independent analysis of the EGRET data [6] indicates a point source whose position is different from the GC at a confidence level beyond 99.9 %.

At energies above 200 GeV, the GC has been observed by VERITAS, CANGAROO and HESS [7, 8, 9]. The spectra as measured by these experiments show substantial differences. This might be due to different integration areas of the signal, a source variability at a time-scale of about one year or inter-calibration problems. Due to the observations under large zenith angles MAGIC could extend the spectrum to higher energies. This is particularly interesting as a cut-off of the spectrum is expected in particle dark matter annihilation scenarios.

## 2. Observations

The Major Atmospheric Imaging Cherenkov telescope (MAGIC [13]) is the largest Imaging Air Cherenkov Telescope (IACT). Located on the Canary Island La Palma at 2200m a.s.l, the telescope has a 17m diameter high reflectivity tessellated parabolic mirror, mounted on a light weight carbon fiber frame. It is equipped with a high efficiency 576-pixel photomultiplier camera, whose analogue signals are transported via optical fibers to the trigger electronics and the 300 MHz FADC readout. Its physics program comprises, among other topics, pulsars, supernova remnants, active galactic nuclei, micro-quasars, gamma-ray bursts and Dark Matter.

The GC culminates at about 58 deg ZA in La Palma. Below 60 deg ZA, it is visible between May and August for about 150 moon-less hours. The GC region has a quite high and non-uniform level of background light from the night sky. This together with the large ZA requires either observations in the false-source tracking (wobble) mode [17] or to take dedicated OFF data.

Within a distance of $1°$ from the GC there are no stars brighter than mag = 8.4, and there are 16 stars with $8 <$ mag $< 9$. At distances between $1°$ and $1.75°$ from the GC the total number of stars with $4 <$ mag $< 9$ is 26. The brightest ones are Sgr 3 (mag = 4.5), GSC 6836-0644 (mag = 6.4) and GSC 6839-0196 (mag = 7.2).

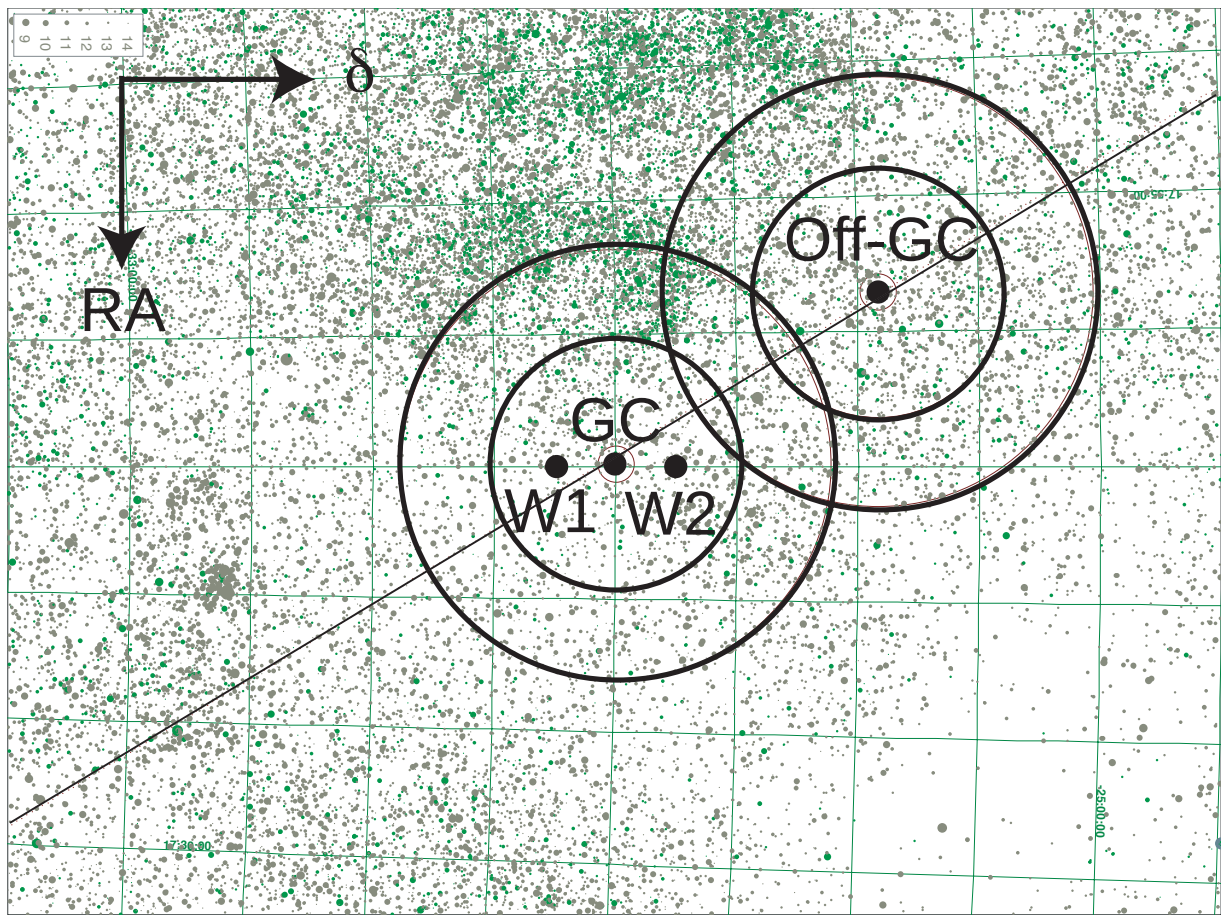

**Figure 1.** Star field around the GC. Stars up to a magnitude of 14 are plotted. The 2 sets of big circles correspond to distances of $1°$ and $1.75°$ from the GC and OFF-GC, respectively. The wobble positions WGC1 and WGC2 are given by the full circles. The $y$ axis is pointing into the direction of decreasing RA, the $x$ axis into the direction of increasing declination. The grid spacing in the declination is 1 degree.

The star field around the GC is quite non-uniform. In the region RA $>$ RA$_{GC}$ $+$ $(4.7$time min$)$ the star field is brighter. The sky directions (WGC1, WGC2) to be tracked in the wobble mode are chosen such that in the camera the sky field relative to the source position (GC) is similar to the sky field relative to the mirror source position (anti-source position). For this reason the directions for the wobble mode are chosen as WGC1 = (RA$_{GC}$, dec$_{GC}$+0.4°) and WGC2 = (RA$_{GC}$, dec$_{GC}$-0.4°). During one wobble mode data taking, 50% of the data is taken at WGC1 and 50% at WGC2, switching between the 2 directions every 30 minutes.

An appropriate OFF region, with a sky field similar to that of the ON region, is centered at the Galactic Plane and contains the bright star Sgr 3 (at (RA, dec) = $(17^h47^m34^s,\ -27°49'51")$ ) in its outer part. The center of the OFF region has the coordinates GC$_{\mathrm{OFF}}$ = (RA, dec) = $(17^h51^m12^s,\ -26°52'00")$. The difference in RA between the GC and $GC_{OFF}$ corresponds to 1.5 degrees. In the same night OFF data is taken directly before or after the ON observations under the same conditions.

After initial observations in September 2004 the Galactic Center direction has been observed for a total of about 28 hours in May-June 2005. This time is split into 10 hours of wobble mode observations, and 9 hours of ON and 9 hours of OFF observations. The data are currently being analyzed.

## 3. Data Analysis

The observations of the Galactic Center are conducted under large ZA. This implies a higher energy threshold but also a larger effective collection area [14]. Thus the otherwise statistics limited high energy domain above 10 TeV could be accessible.

In our preliminary analysis we used the Random Forest method [12] for the gamma hadron separation. Each event is tagged with the parameter hadronness $\in [0;1]$ which is a measure for the probability to be a background event. The smaller the hadronness the higher the probability to be a gamma event and the larger the hadronnes the higher the chance to be a background event.

To train the Random Forest, high ZA (60° ZA) Monte Carlo (MC) gamma showers were generated with energies between 200 and 30,000 GeV. The differential spectral index of the generated spectrum is $-2.6$, conforming with the energy spectrum of the Crab nebula. The MC sample was divided into two sub-samples. The Random Forest was trained with one sub-sample of the MC gammas and a sub-sample of the GC-OFF data which represents the background events. Thereafter the trained Forest was tested with the other MC sub-sample and a different sample of OFF data. As training parameters we used the Hillas parameters SIZE, DIST, WIDTH, LENGTH, CONC, and M3Long [16]. The training was done for SIZE $>$ 200 photo electrons.

In order to develop and verify the MAGIC analysis at high zenith angles Crab data in the interesting ZA range around 60° have been taken in January 2005. The gamma energy spectrum of this data can be reconstructed and it is consitent with existing measurements. Figure 2a) shows a sky map of gamma ray candidate excess events from direction of the Crab nebula with a lower cut on SIZE of 600 photo electrons.

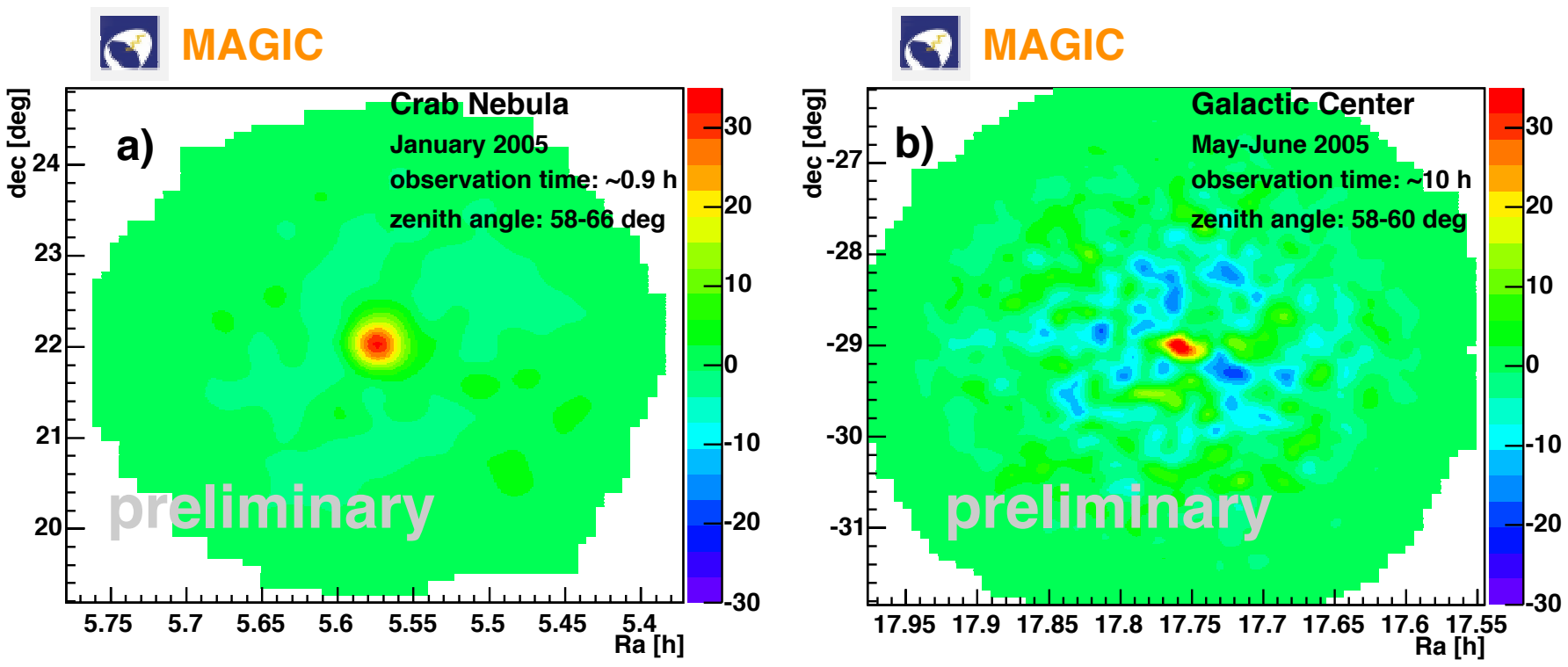

**Figure 2.** Sky maps of candidate gamma ray excess events in the directions of the Crab nebula (a) and the Galactic Center (b) with lower Size cuts of 600 photo electrons (corresponding to 2 TeV) and a hadronness$< 0.15$.

Figure 2b) shows a sky map of gamma ray candidate excess events from the direction of the Galactic Center with a lower SIZE cut of 600 photo electrons and a hadronness $< 0.15$. This size cut corresponds to an energy threshold of about 2 TeV. Although not yet significant, the observed excess in the direction of the Galactic Center is compatible within errors with previous measurements [9, 7].

## 4. Discussion

The GC is an interesting target in all wavelength bands. First detections of the GC by the Veritas, Cangaroo and HESS collaborations were made. The measured fluxes exhibit significant differences. These may be explained by calibration problems, by time variations of the source or by different integration areas due to different point spread functions. The nature of the source of the VHE gamma rays has not yet been identified.

Conventional acceleration mechanisms for the VHE gamma radiation utilize the accretion onto the black hole and the diffusive shock of supernova remnants. The GC is also expected to be the brightest source of VHE gammas from particle dark matter annihilation [10, 11, 15]. Most probably the main part of the observed gamma radiation is not due to dark matter annihilation [3]. Nevertheless it is interesting to investigate and characterize the observed gamma radiation as a contribution due to dark matter annihilation is not excluded.

Data taken by MAGIC will help to determine the nature of the source and to understand the flux discrepancies. Due to the large zenith angles, MAGIC will have a large energy threshold but also a large collection area and good statistics at the highest energies. The measurements may also be used to inter-calibrate the different IACTs.

## A. Acknowledgements

The authors thank A. Moralejo for helpful discussions about the Monte Carlo simulations. The project MAGIC is supported by MPG (Max-Planck-Society in Germany), and BMBF (Federal Ministry of Education and Research in Germany), INFN (Italy), IFAE (Spain) and IAC (Instituto de Astrophysica de Canarias).

# Search for pulsed VHE Gamma ray emission from the Crab Pulsar

M. López[a], N. Otte[b], M. Rissi[c], P. Majumdar[b], J.A. Barrio[a], M. Camara[a], J.L Contreras[a], R. de los Reyes[a], M.V. Fonseca[a], O.C. de Jager[e], O. Kalekin[d], F. Lucarelli[a], E. Oña-Wilhemi[f], I. Oya[a], for the MAGIC collaboration

*(a) Fac. C.C. Físicas, Universidad Complutense de Madrid, Avd. Ciudad Universitaria S/N.20840, Madrid, Spain*
*(b) Max-Planck-Institut für Physik, Föhringer Ring 6, D-80805 München, Germany*
*(c) Institute for Particle Physics, Swiss Federal Institute of Technology (ETH) Zurich, Switzerland*
*(d) Institut für Physik, Humboldt-Universität Berlin, Germnay*
*(e) Space Research Unit, Northwest University, Potchefstroom, 2520, South Africa*
*(f) Instituto de Física de Altas Energías, Univeridad Autónoma de Barcelona, 08193, Bellatera, Spain.*

Presenter: M. López (marcos@gae.ucm.es)

Pulsed gamma-ray emission has been observed from the Crab pulsar by the EGRET instrument up to energies of 5 GeV. With the 17 meter MAGIC Telescope we have searched for pulsed gamma rays from Crab, and obtained a flux upper limit for its pulsed emission.

## 1. Introduction

The Crab pulsar (rotation period $33\,\mathrm{ms}$) and its nebula are one of the most studied objects in VHE gamma-ray astronomy. The pulsar itself is emitting over over a broad energy range from radio up to gamma-rays. It is also the only known pulsar which presents the same light curve at all energies. The most energetic pulsed $\gamma$–ray emission detected is at 5 GeV and has been measured with the EGRET instrument on board of CGRO [1].

The most popular models describing the pulsed emission are the *polar cap* [2] and the *outer gap* [3] models. Both predict a sharp energy cutoff in the emission spectrum between a few GeV and few tens of GeV. In the polar cap scenario, electrons are accelerated above the pulsar polar cap radiating $\gamma$-rays via curvature and synchron radiation. Due to the strong magnetic fields of the pulsars magnetosphere ($\sim 10^{12}$ Gauss) the $\gamma$-rays can undergo magnetic pair production. As a consequence of this in the polar cap model a super–exponential cutoff is expected in the $\gamma$-ray flux spectrum above a cutoff energy $E_c$. In the outer gap model $\gamma$-rays are emitted close to the light cylinder of the pulsar. In this model the cutoff is determined by photon-photon pair production which has a weaker energy dependence than magnetic pair production. Therefore a higher cutoff energy is expected.

By precisely determining the cutoff energy of the spectrum one would be able to discriminate between several models thus answering the open question about the acceleration site of electrons in the magnetosphere of the pulsar. Up to now no instrument has been able to detect pulsed emission beyond a few GeV. One possible way to detect $\gamma$-rays in the energy region of interest is to use the imaging air Cherenkov technique. However, until some years ago air Cherenkov telescopes had a too high energy threshold. MAGIC [4] is the largest air Cherenkov telescope with a reflector diameter of 17m. The current trigger threshold of MAGIC is about 50 GeV, well suited to perform pulsar studies. The central pixel of the MAGIC camera is used to detect the optical emission from the Crab Pulsar and other objects to perform correlation studies [5]. The commissioning phase of the telescope finished last summer.

## 2. Data Analysis

### 2.1 The data sample

The data analyzed here were taking during the first Crab campaign of the MAGIC telescope after its commissioning phase. The data were taken with two different trigger conditions, namely, in the standard trigger mode, and with a special trigger condition. In the later, the trigger region is restricted to a narrow ring of the inner MAGIC camera, since this configuration provides a lower trigger threshold, as well as a better hadron suppression at the trigger level. Table 1 summarizes the two data sample used in this analysis.

**Table 1.** Data sample used in this analysis.

| Mode | Date | $T_{obs}$ $(hours)$ |
|---|---|---|
| Standard trigger | Sep. 2004: 21, 22 | 0.8 |
| | Oct. 2004: 10,11,22 | 2.4 |
| Special trigger | Oct. 2004: 16, 21, 23 | 4.2 |
| | Nov. 2004: 19, 21 | 3.1 |

The standard calibration of the MAGIC telescope uses a set of light pulse generators which illuminate the PMT camera uniformly. This procedure provides an absolute calibration of the camera and the signal processing chain of the telescope. The analog signals from the PMTs of the camera are digitized by 300 MHz Flash-ADCs. The conversion factor from ADC counts to photoelectrons is obtaining by means of the so called F-Factor method. Once the data has been calibrated, a cleaning algorithm is applied to the shower images in order to remove the noise of the night sky background. The final images are then parameterized in terms of the Hillas parameters [6].

### 2.2 Timing analysis

For each event which triggers the telescope the exact arrival time is recorded using a Rubidium clock synchronized with a Global Positioning System (GPS) at the beginning of each second, which provide a precision of 200 ns. For the timing analysis, all the arrival times are transformed to the solar system barycenter, using the JPL DE200 planetary ephemerides and the TEMPO package [7]. The corrected times were folded modulo the pulse period to obtain the corresponding rotational phases of the events, according to a Taylor expansion around the known Crab radio ephemeris. The ephemeris used for each observational period were obtained from the Jodrell Bank public web page [8]. In order to test our analysis software, we used the optical data taken with the *Central Pixel* of the MAGIC telescope (for a detailed description see [5] in these proceedings).

#### 2.2.1 *Periodicity search*

Standard gamma/hadron separation methods based on cuts on the distribution of the Hillas parameters describing the images, remove a large fraction of events a low energies, and, therefore, they are not suitable for pulsar searches, specially for the very low energy cutoff expected for the Crab pulsar. Since the signal, if at all observable, is expected at the very trigger threshold of the MAGIC telescope, we applied a cut in the size of the images of 300 photons (equivalent to $\sim 55$ photoelectrons) thus removing all the high energy showers and retaining only the low energy events. This cut reduces the background rate by 93% while losing only 34% of the gammas (assuming a super-exponential cutoff for the Crab pulsar nebula at 20 GeV, according to [10]).

Then, to search for the presence of a periodic signal at the Crab frequency, we applied different uniformity tests to the whole data sample surviving this cut . We also did a blind periodicity search around the expected Crab frequency. In either case, no evidence of a significant signal was found. The obtained light curve can be seen in Figure 1, which is consistent with an uniform distribution.

#### 2.2.2 *Flux upper limits*

Since there was no evidence of the pulsed signal in our data, we proceded to calculate the upper limit on the pulsed Crab emission. For obtaining conservative upper limits, only the data taken in the standard trigger mode have been used, using the same gamma/hadron separation method that we employed to extract the unpulsed flux of the Crab nebula [11], but just with looser cuts. This made the energy threshold higher but easier to compare with the DC component. We have derived the upper limits from the H-test. The 3 sigma upper limit is parameterized in terms of the value $H$ obtained in the uniformity test and the expected duty cycle of the pulse profile $\delta$, as [9]:

$$x_{3\sigma} = (1.5 + 10.7\delta) \cdot (0.174\, H)^{(0.17+0.14\delta)} \cdot exp\big((0.08 + 0.15\delta) \cdot [log_{10}(0.174 \cdot H)]^2\big) \quad (1)$$

where $x = p\sqrt{N}$ is $p$ the pulsed fraction and $N$ the total number of recorded events. The pulse profile of the Crab pulsar above few GeVs remains unknown. For this analysis we assumed that it retains the same shape as the one measured by EGRET, i.e., we assumed a double peak pulse profile with a duty cycle of 21%. The integral upper limits obtained are reported in table 2. They have been calculated for two different lower SIZE cuts, resulting in an analysis energy threshold of 90 and 150 GeV respectively. The upper limits are compared to previous results by other experiments in Figure 1. In order to constraint the spectral energy cutoff, we have extended the EGRET power law [14] to our energy domain by assuming an exponential cutoff. The higher cutoff compatible with our upper limit is $E_c \leq 60$ GeV.

**Table 2.** Flux upper limits derived at different threshold energies $E$. $H$ is the value obtained in the H-test, and $p$ and $N$ are the upper limit to the pulsed fraction and the total number of events after cuts respectively.

| E (GeV) | H | x=p$\sqrt{N}$ | F$_{ul}$ ($cm^{-2}s^{-1}$)x10$^{-10}$ |
|---|---|---|---|
| $\geq 90$ | 5.7 | 3.6 | 2.0 |
| $\geq 150$ | 4.2 | 3.4 | 1.1 |

## 3. Conclusions

Data taken by the MAGIC telescope during last winter has been used to search for pulsed $\gamma$-ray from the Crab pulsar. We found no evidence for its pulsed emission in our analysis at the expected radio frequency. We have calculated upper limits at different energies for a 3 $\sigma$ confidence level. Assuming a exponential cutoff for the Crab spectrum we conclude that this cutoff energy must be below 60 GeV. The upper limits derived here, do not allow yet to discriminate between the outer gap and polar cap models.

Apart from the Crab, other pulsars have been observed by MAGIC, which analysis can be found elsewhere in these proceedings [16].

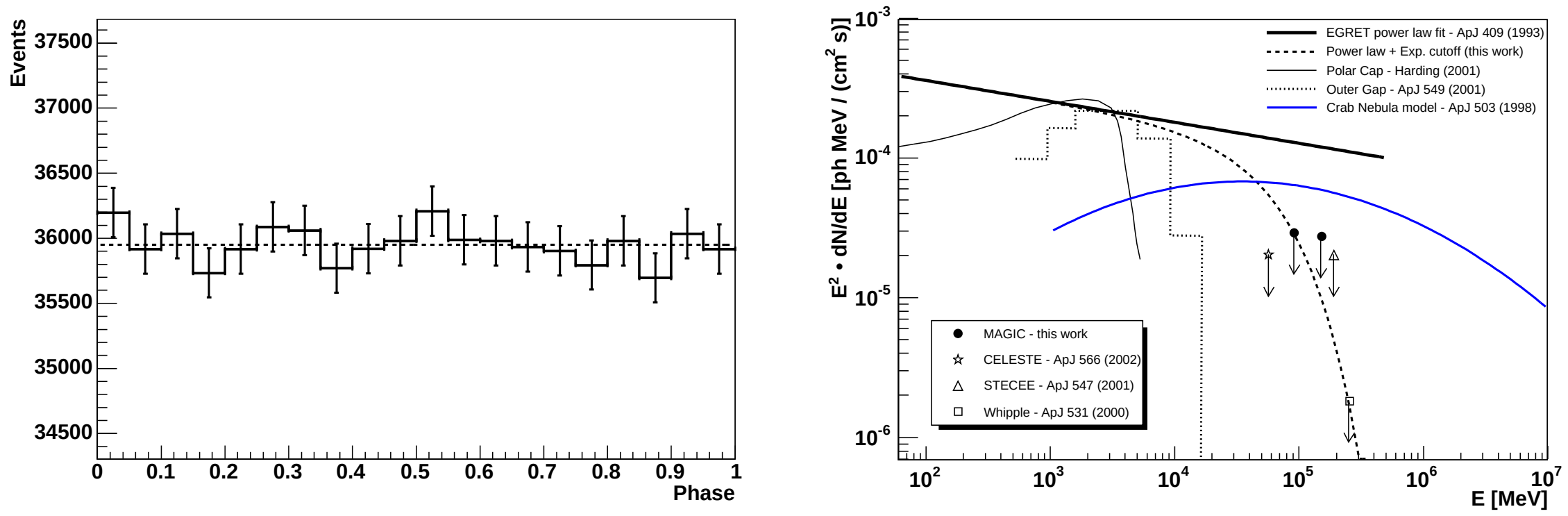

**Figure 1.** Left: Light curve at the expected Crab radio frequency selecting only those events below a SIZE cut of $\sim 55$ photoelectrons, i.e., close to the telescope energy threshold, and without applying any additional gamma/hadron separation cut. Right: Pulsed photon spectrum of the Crab pulsar. The full circles represents the differential flux upper limits obtained in this work at different energy thresholds. For comparison, the upper limits obtained by other experiments are also shown (see e.g. [12] and [13]). The solid line is the power law fi t to the EGRET data [14], whereas the blue line is a model for the crab nebula spectrum [15]. The dashed line represents the extension of the EGRET power law into the MAGIC energy domain with an exponential cutoff constraint by our upper limit.

## 4. Acknowledgments

The MAGIC project is supported by the Spanish CICYT agency (under project No. FPA2003-9543-C02-01), the Max-Planck-Society (Germany), the BMBF (Federal Ministry of Education and Research in Germany), the INFN (Italy) and the IAC (Instituto de Astrophysica de Canarias).

# Observations of Mrk 421 with the MAGIC Telescope

D. Mazin[a], J. Flix[c], F. Goebel[a], E. Lindfors[b], J. Lopez Munoz[c], M. Lopez Moya[d], A. Moralejo[e], V. Scalzotto[e], T. Schweizer[f], and R.M. Wagner[a]
for the MAGIC Collaboration
*(a) Max-Planck-Institut für Physik, Föhringer Ring 6, 80805 München, Germany*
*(b) Tuorla Observatory, Väisäläntie 20, 21500 Piikkiö, Finland*
*(c) Institut de Fisica d'Altes Energies, 08193 Bellaterra (Barcelona), Spain*
*(d) Universidad Complutense de Madrid, Facultad de Ciencias Fisicas, 28040 Madrid, Spain*
*(e) Dipartimento di Fisica, Universita di Padova and INFN sez. di Padova, Via Marzolo 8, 35131 Padova, Italy*
*(f) Humboldt Universität zu Berlin, Institut für Physik, Newtonstraße 15, 12489 Berlin, Germany*
Presenter: D. Mazin (mazin@mppmu.mpg.de), ger-mazin-D-abs1-og23-Oral

The AGN Mrk 421 was observed during moderately high flux states between November 2004 and April 2005 with the MAGIC telescope shortly after the end of its commissioning phase. Here we present a combined analysis of a large data sample recorded under different observational conditions. The integrated flux level is observed to vary by more than a factor of 2 on different time scales. The energy spectrum between 100 GeV and 2 TeV is well described by a power-law with a photon index $\Gamma = 2.6$ independent of the flux level.

## 1. Introduction

The 17 m diameter MAGIC telescope [1], located on the Canary Island La Palma (2200 m a.s.l.) has completed its commissioning phase in September 2004. The main design goal was to explore the low energy range, eventually down to 30 GeV. The first physics observations in Winter 2004/05 and in Spring 2005 included observations of the well established TeV blazar Mrk 421. In total, 19 nights of data have been taken on this source, the observation times per night ranging from 30 minutes to 4 hours.

Mrk 421 is the closest known TeV blazar (redshift $z$ = 0.031). It was the first extragalactic $\gamma$-ray source discovered in the TeV energy range using Imaging Air Cherenkov Telescopes [2, 3]. Mrk 421 is the source with the fastest observed flux variations reported among TeV $\gamma$-ray emitters, with variations up to one order of magnitude and occasional flux doubling times as short as 15 min [4]. Variations in the hardness of the TeV $\gamma$-ray spectrum during flares have been reported in [5, 6]. Simultaneous observations in the X-ray and GeV-TeV bands show a significant correlation of the fluxes. This supports the hypothesis that the same population of electrons generates X-ray photons via synchrotron radiation and GeV-TeV photons via inverse-Compton scattering.

## 2. Data sample

The data taken on Mrk 421 between November 2004 and April 2005 was divided into 4 samples. Due to changes in the hardware the data before and after February 2005 were treated separately. Most of the data was taken at small zenith angles ($ZA < 30°$), resulting in a low analysis energy threshold. Only 1.5 h in December 2005 were taken at $42° < ZA < 55°$ during simultaneous observations with the H.E.S.S. telescope system [7]. The standard operation mode for MAGIC is the ON-observation, where the source is in the center of the camera. In April 2005 part of the data was taken in wobble mode. In this mode the source is tracked alternately in two opposite positions 0.4° off the center. The definition of the 4 data samples is summarized in

**Table 1.** Results of the Mrk 421 data. Samples I+II were recorded in November 2004 - January 2005, while Samples III+IV were taken in April 2005

| sample | on time | zenith [°] | mode | $E_{thr}$[GeV] | $N_{on}$ | $N_{off}$ | $N_{excess}$ | sigma |
|---|---|---|---|---|---|---|---|---|
| I | 4.70 h | 9.3 - 31.2 | ON | 100 | 7458 | 5084.0 ± 59.3 | 2374.0 ± 102.1 | 23.26 |
| II | 1.41 h | 42.4 - 55.0 | ON | 300 | 593 | 315.9 ± 14.9 | 277.1 ± 27.3 | 10.13 |
| III | 7.88 h | 9.2 - 27.5 | ON | 100 | 8116 | 5089.8 ± 59.4 | 3026.2 ± 104.5 | 28.96 |
| IV | 9.57 h | 9.4 - 32.4 | wobble | 100 | 9296 | 5668.3 ± 45.1 | 3627.7 ± 98.3 | 36.89 |

Table 1. For each data sample a separate Monte-Carlo sample was simulated taking into account the particular observation conditions.

The full data set consists of 29.0 hours. Runs with problems in the hardware or unusual trigger rates have been rejected in order to ensure a stable performance and good atmospheric conditions. The total observation time amounts to 23.2 h after cuts.

## 3. Analysis

For calibration, image cleaning, cut optimization and energy reconstruction the standard analysis techniques of the MAGIC telescope were applied (see [9]). For the $\gamma$/hadron separation a multidimensional classification technique based on the Random Forest method [10] was used, with classical Hillas parameters [11] like $Width$, $Length$ and $Size$ but excluding $Alpha$ as input parameters.

For the ON-mode data the remaining background was estimated by performing a second order polynomial fit to the $Alpha$ distribution (see Fig. 1) in the range between 30° and 90°, where no $\gamma$ events were expected. The signal was then determined as the number of observed events in the range $Alpha < \alpha_0$ exceeding the fit extrapolated to small $Alpha$, where $\alpha_0$ is energy dependent and has a typical value of 12°. For the wobble-mode data the image parameters were calculated both with respect to the position of the source (ON) and with respect to an OFF position on the opposite side of the camera. The background was then estimated comparing the two $Alpha$ distributions. In order to avoid an unwanted contribution of $\gamma$-events in the OFF sample an anti$Alpha$ cut was applied viz. $\gamma$-candidates (i.e. $Alpha < \alpha_0$) from the ON sample were excluded from the OFF sample and vice versa.

The energy spectra were obtained using spillover correction factors to compensate for instrumental effects. This analysis aims at producing solid results above 100 GeV. The lower energy regime requires additional studies especially concerning the background rejection.

## 4. Results

During the whole observation period Mrk 421 was found to be in a moderate to high flux state resulting in clear signals in all four data samples. Fig. 1 shows the $Alpha$ distribution of the $\gamma$-candidates of the combined samples I and III with an energy threshold of 200 GeV. An excess of 2100 events was found, which corresponds, for the given background, to more than 40 standard deviations. The number of excess events and the significances for the individual samples are summarized in Table 1. Fig. 2 shows a sky map produced with the DISP method [12] using sample IV after applying the same cuts as for the $Alpha$ distribution.

The integral fluxes above 300 GeV averaged over each night of observation are shown in the upper part of

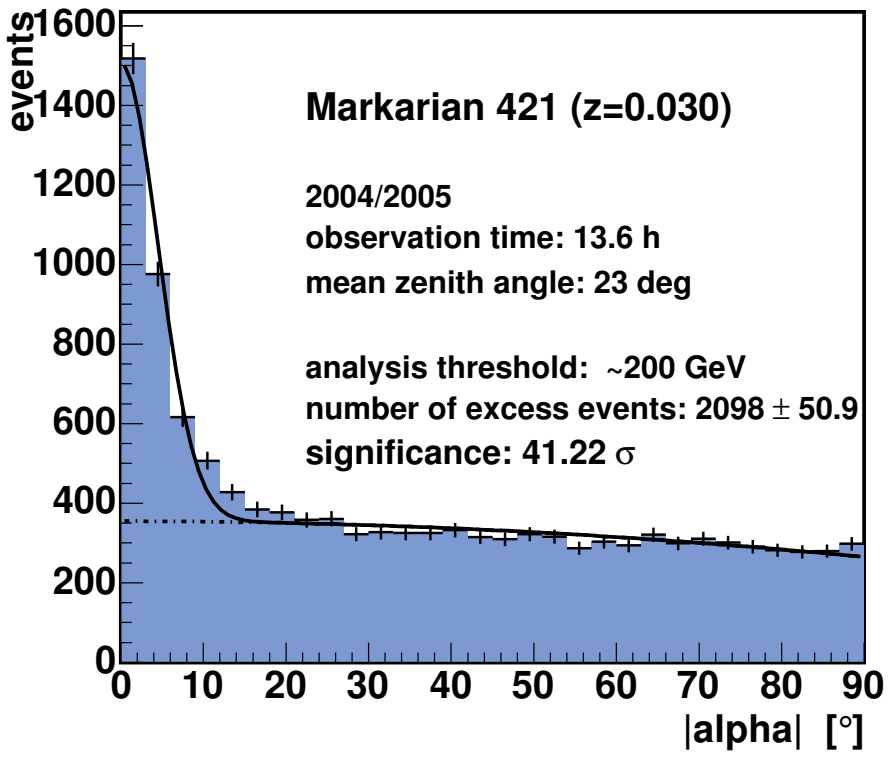

**Figure 1.** *Alpha* distribution for the combined data samples I+II+III with $E > 200$GeV.

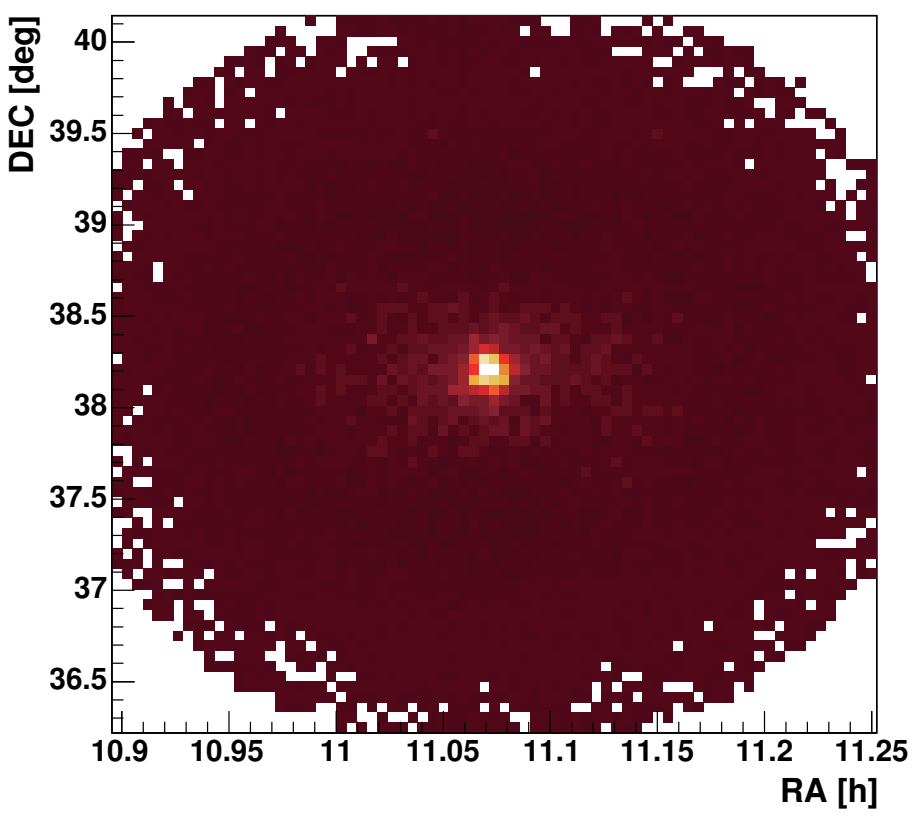

**Figure 2.** Sky map of Mrk 421 using the DISP method [12].

Fig. 3. Significant variations of up to a factor of 3 can be seen. In order to avoid systematic effects resulting from different observation conditions the light-curve was produced using only samples I and III. The relatively high energy threshold of 300 GeV ensures that the results are independent of the actual thresholds during each night. In the lower part of Fig. 3 the corresponding flux in the X-ray band as observed by the ASM detector [8] on-board the RXTE satellite are shown for the analyzed observation period. Particularly for the April data a clear correlation between the X-ray and the GeV-TeV $\gamma$-ray activity can be observed.

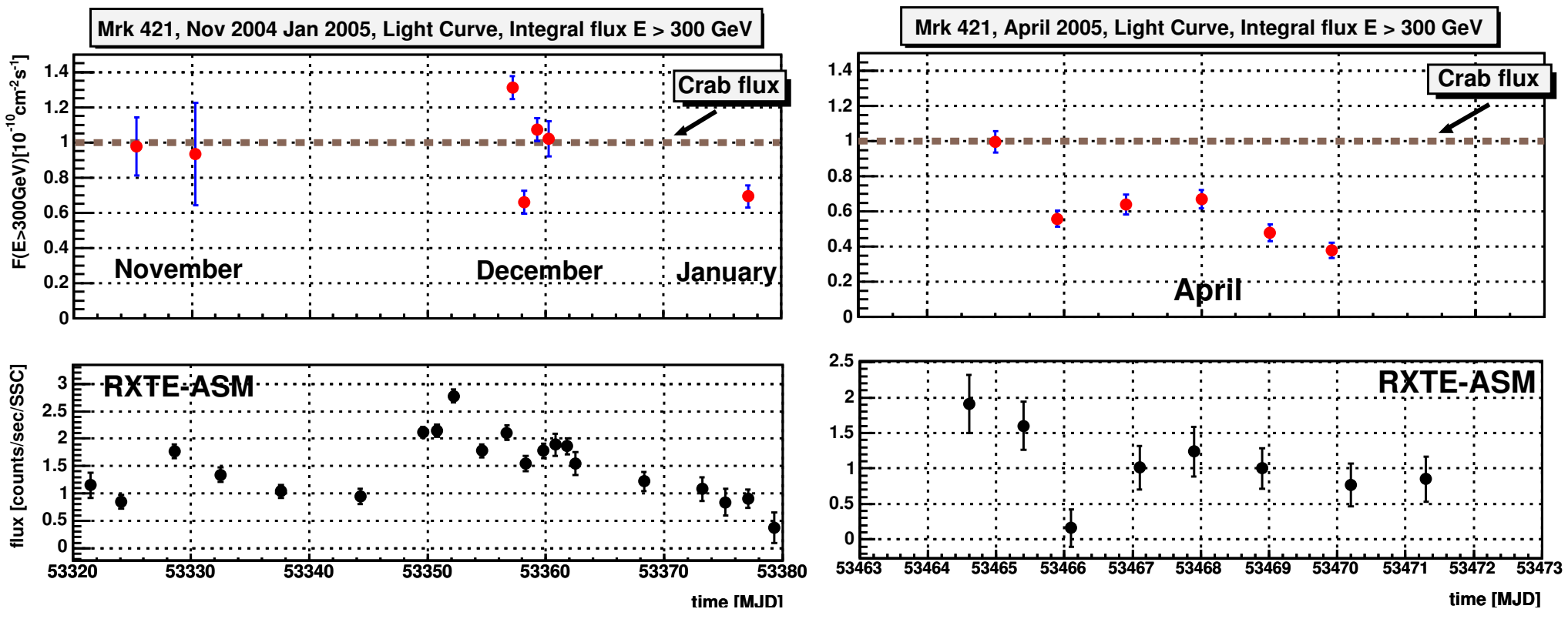

**Figure 3.** Light curve of Mrk 421 between November 2004 and April 2005 above 300 GeV (upper plots) and the corresponding X-ray flux as observed by ASM.

For the day of highest activity in April the intra-night variability is shown in Fig. 4 in bins of 20 minutes. In order to extend the time coverage as much as possible data sample III (before midnight) and sample IV (after midnight) have been combined. The observation suggests doubling times of less than 1 h.

For the spectrum calculation the data taken between December 2004 and April 2005 (samples I+III only) have been divided into a high and a low flux state (highest 4 and lowest 7 nights in Fig. 3). The corresponding energy spectra (see Fig. 5) have been fitted with a power-law ($\mathrm{d}N/\mathrm{d}E = N_0 \cdot (E/\mathrm{TeV})^{-\Gamma}$) in the energy range between 100 GeV and 2 TeV yielding $\Gamma = 2.61 \pm 0.05(stat.)$ for the low flux sample in good agreement with $\Gamma = 2.65 \pm 0.06(stat.)$ for the high flux sample.

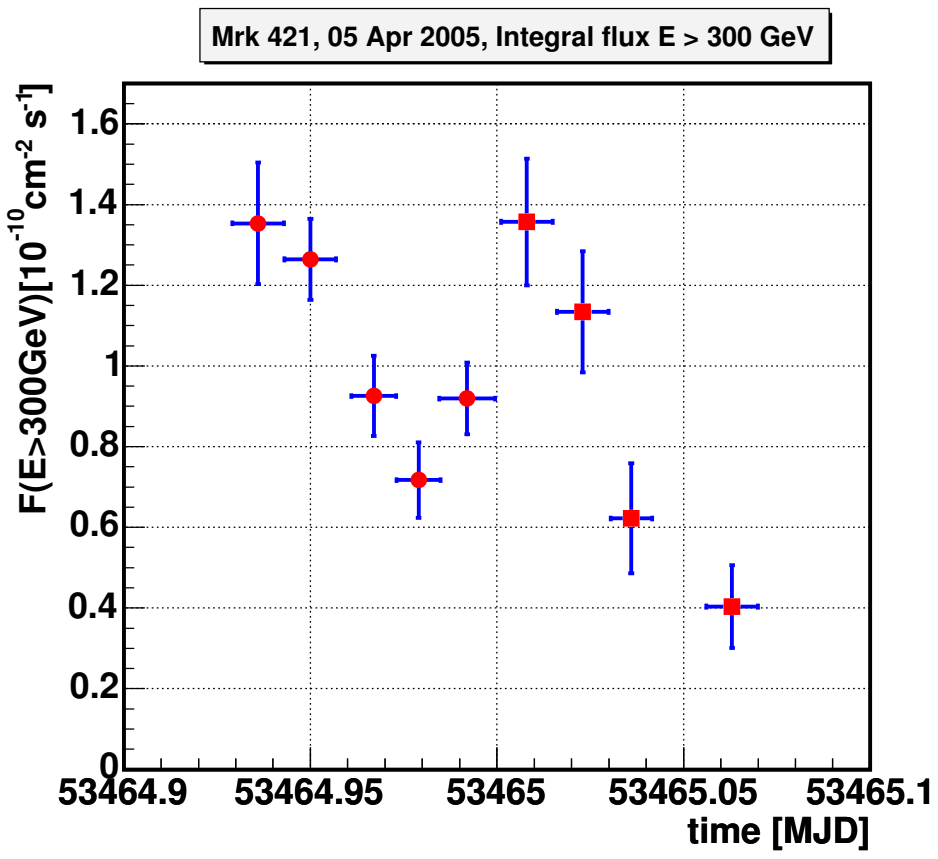

**Figure 4.** Light-curve for the night MJD 53465 in bins of 20 minutes (sample III: circles, sample IV: squares)

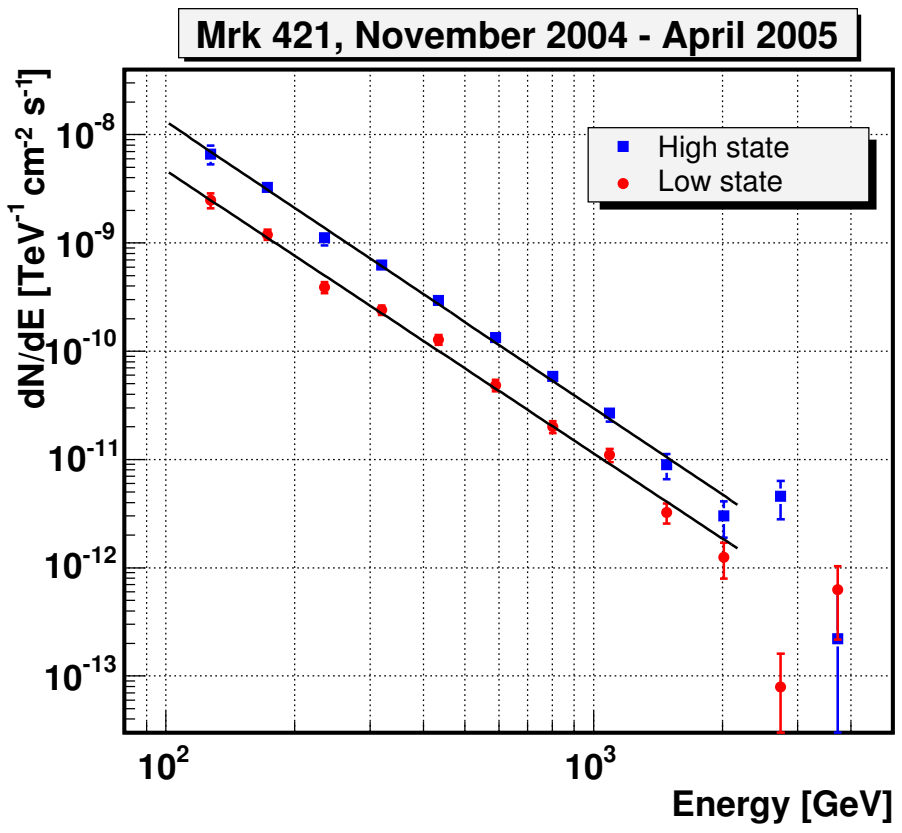

**Figure 5.** Differential energy distribution for Mrk 421 for high and low flux samples as described in the text.

## 5. Conclusion

Mrk 421 has been observed with the MAGIC telescope during several months after its commissioning phase. The data have been used to produce the first energy spectrum of this source extending down to 100 GeV. Both for high and low flux state the spectra are well described by a power law with photon index $\Gamma = 2.6$. The light-curves above 300 GeV show clear correlation with the X-ray fluxes. On MJD 53465 flux variations of a factor of 2 are observed on an hourly time scale.

*Acknowledgments:* We would like to thank the IAC for excellent working conditions. The support of the German BMBF and MPG, the Italian INFN and the Spanish CICYT is gratefully acknowledged.

# First pulsar observations with the MAGIC telescope

E. Oña-Wilhelmi[a], R. de los Reyes[b], J.L. Contreras[b], C. Baixeras[c], J.A. Barrio[b], M. Camara[b], J. Cortina[a], O.C. de Jager[d], M.V. Fonseca[b], M. L$o$pez[b], I. Oya[c] and J.Rico[a]
for the MAGIC collaboration.
*(a) Institut de Fisica d'Altes Energies (IFAE), Universidad Autonoma de Barcelona, 08193, Bellaterra, Spain.*
*(b) Dpto. Fisica Atomic, Nuclear y Molecular, Universidad Complutense de Madrid, 28040, Madrid, Spain*
*(c) Dpto. Fisica, Universidad Autonoma de Barcelona, 08193, Bellaterra, Spain.*
*(e) Space Research Unit, Northwest University, Potchefstroom 2520, South Africa*

Presenter: E. Oña-Wilhelmi (emma@ifae.es), spa-ona-wilhelmi-E-abs1-og22-oral

A few regions of the sky containing pulsars have been observed by the MAGIC Telescope [2] during its commissioning phase, namely PSR B1957+20, and PSR J0218+4232. In this work we report on the analysis of these data, looking for $\gamma$-ray emissions both in continuous and pulsed mode. Constrains to different theoretical models about $\gamma$-ray emission from pulsars and plerions will be discussed.

## 1. Introduction

During the first year of operation of the MAGIC telescope quite some time has been devoted to the Crab pulsar and a dedicated contribution is presented in this conference [16]. Beside Crab, for the current study, a small subset of pulsar systems was observed and analyzed as part of an ongoing program to search for galactic sources of GeV emission. These sources were chosen based on their similarity to some other detected sources and on the prediction of emission given the various models of $\gamma$-ray production in pulsars. A deeper analysis have been done with two of these sources, PSR B1957+20 and PSR J0218+4232.

No evidence has been found of pulsed or unpulsed emission from the pulsars at high energies. These non-detections place limits on photon densities, magnetic field strengths, and emission regions in plerionic systems, since both pulsars are in binary systems. Both, PSR B1957+20 and PSR J0218+4232 are millisecond pulsars in energetic binary systems, detected in X-rays. The X-ray emission may be due to the interaction of the pulsar wind with the companion. Those energetic systems are believed to be as well $\gamma$-ray emitters, not only by interaction between the two stars [15][7] but also, several models of millisecond pulsars [4] propose pulsed $\gamma$-ray emission from the pulsar inner magnetosphere up to $\sim$100 GeV. Different modelization of millisecond pulsars ([13] (and reference therein) and [6]) predict a high cutoff energy $E_o$ in their spectra, at a $\sim$ 100 GeV, while for canonical pulsars this cutoff is expected to happen at a few GeV. These models are based in magnetic field strength and pair production considerations: although these pulsars have low surface magnetic fields ($B_s \sim 10^9\ G$), their short periods allow them to have large magnetospheric potential drops, but the majority do not produce sufficient pairs to completely screen the accelerating electric field. Thus the spectra are very hard power-laws with exponential cutoffs up to 100 GeV. On the other hand, ablation and heating of the companion star are believed to be caused by X- or $\gamma$-rays generated in a intrabinary shock between the pulsar wind and the star. This process has been observed by HESS in the similar system PSR B1259-63. The observed flux of $\gamma$-rays, modulated within the orbit, were detected at 5% of the Crab emission.

The observations of each of these pulsars were carried out in very different conditions and therefore it is necessary to analyze them in different ways to extract a steady signal. The different procedures and observation conditions will be explained for each pulsar. In addition statistical tests were applied to the $\gamma$-like events arrival times to search for pulsed emission. The standard corrections were performed in arrival times and the ephemeris used were obtained from the ATNF database, summarized in Table1.

| | PSR J0218+4232 | PSR B1957+20 |
|---|---|---|
| Epoch (MMJD) | 50864.00 | 48196.00 |
| Puls. Frequency (Hz) | 430.4610663457 | 622.122030511 |
| Freq. Deriv. s/s | -1.4340E-14 | -6.5221E-15 |
| Puls. Period (s) | 0.0023230904678309 | 0.00160740168480632 |
| Rot Period (days) | 2.028846084 | 0.3819666069 |

**Table 1.** *PSR J0218+4232 & PSR B1957+20 ephemeris (ATNF pulsar database).*

## 2. Observations and Data Analysis

### 2.1 PSR B1957+20

PSR B1957+20 is one of the fastest pulsars known, with period P∼1.6 ms and spin down luminosity $\dot{E} = 1 \times 10^{35}$ erg$\cdot s^{-1}$. The pulsar is in a 9.16-hour binary orbit with a low-mass companion star. Observations of the pulsar PSR B1957+20 were carried out with MAGIC in October 2004 ($6^{th}$-$16^{th}$) for 6 hours effective ON time, at low zenith angle ($<30^o$). The observations were done in a standard ON-OFF mode, pointing the telescope to the pulsar position.

The total data set has been analyzed to look for emission in both steady and pulsed modes. Only high quality data were considered for the analysis. After calibration and image cleaning, Cherenkov showers were reconstructed using the standard Hillas parameter technique [8]. To reduce the hadronic background, optimum cuts on the Hillas parameters were applied to select $\gamma$-like events. This optimization was done with the Random Forest algorithm [3], which was trained to recognize $\gamma$-ray events with Crab Nebula ON data from the same epoch and same zenith angle.

For the pulsed emission analysis, the arrival times of the Cherenkov events were registered by a GPS clock. All arrival times were then transformed to the solar system barycenter using the JPL DE200 planetary ephemerides and folded modulo the period relevant to the epoch and the source. Some very loose cuts were applied since the signal is expected to be dominant at low energies where the Hillas technique fails to discriminate $\gamma$-rays from the hadronic background.

Once the necessary corrections are applied, which also take into account the pulsar motion with respect to the companion, several periodicity tests were performed: not only $\chi^2$ to test against a uniform distribution, but also tests based on Fourier transformations which are able to resolve more sophisticated pulse shapes ($Z^4_m$ and H test) [5].

### 2.2 PSR J0218+4232

PSR J0218+4232 was discovered in 1995 [11] as a luminous radio pulsar with pulse period of 2.3 ms. It orbits a degenerate companion with orbital period of about 2 days. The radio profile of the pulsed emission is complex, showing three peaks and a continuous (DC) component.

This millisecond pulsar has also been detected in soft and hard X-rays, and belongs to the reduced club of those showing X-ray modulation at the rotational period. As an X-ray pulsar it has a high luminosity, hard power-law shaped X-ray spectrum and a profile composed of narrow pulses. The X-ray emission is consistent with a structure of non-thermal hard pulses superimposed on a continuum of softer spectrum. The source is punctual with a size below 1 arc second.

In addition of the above properties this pulsar may represent the only detection of a millisecond pulsar in $\gamma$-

| PULSAR | F ($> 115\, GeV$) ($\times 10^{-12}\, cm^{-2} s^{-1}$) |
|---|---|
| PSR B1957+20 | 4.5 |
| PSR J0218+4232 | 1.3 |

**Table 2.** *PSR J0218+4232 & PSR B1957+20 flux upper limits for the steady emission.*

rays. It is positionally coincident with the EGRET source 3EG J0222+4253 which has been related with both the AGN 3C66A and this source. Kuiper et al.[9] reported a 3.5 $\sigma$ detection of a pulsed signal in EGRET data whose significance increased later on to 4.9 $\sigma$ when it was shown that the peaks in the $\gamma$ profile were aligned with the X-ray peaks.

The analysis of PSR J0218+4232 was a byproduct of dedicated observations on the AGN 3C66A, since for MAGIC the pulsar happens to be in the same field of view of the BL Lac object. Therefore the pulsar appears at a distance of 0.973$^{\circ}$ of the center of the telescope camera, close to the limit of the trigger area ($\sim 1^{\circ}$). This fact implies a reduced effective area and MAGIC sensitivity in a factor $\sim$35% [14].

The data analyzed belong to August 2004 ($20^{th}$-$26^{th}$) and October 2004 ($10^{th}$-$21^{th}$) with a total ON observation time of $\sim$ 13 h. The mean observation zenith angle is $\sim$ 15$^{\circ}$. Two different analysis scenarios were tried:

- In the first scenario a lower cut on the number of photoelectrons in the image (*size* parameter) demanding at least a value of 150, was applied. Images were reconstructed from the source offset position. Those high energy/size images thus selected had enough quality to be treated through a standard set of cuts. This analysis has a nominal threshold of about 115 GeV.
- The second approach was to select only low energy events. It allows to greatly reduce the background from cosmic rays events and exploit the soft spectrum that most models predict for the pulsed emission of pulsars. On the other hand no standard imaging is possible in this regime since images consist of very few pixels with small amplitudes.

## 3. Results

No excess is observed in the total data set for the millisecond pulsars neither for the binary systems. The upper limit on the integrated flux above a certain energy $E_t$ was inferred from the MAGIC sensitivity [10] for a point-like source, as obtained from Monte Carlo calculations. Upper limits at $2\sigma$, 95% confidence level, are listed in Table 2 for the steady emission. No evidence of pulsed emission was detected, either from the orbital modulation (in the case of PSR B1957+20) or from the pulsars at the expected frequency (see Fig. 1 and Fig. 2). Upper limits for the pulsed emission using the method explained by O.C de Jager [5] will be presented at the meeting.

## 4. Discussion

The data tabulated in Table 2 summarize the upper limits for unpulsed emission from the 2 systems under study. The energy threshold achieved in this analysis is still high to constrain the emission due to CR (up to 50 GeV) in the theoretical spectra [4] for isolated pulsars. However, the flux upper limit for the binary pulsar PSR B1957+20 is lower than the predicted value of [7]. This discrepancy can perhaps be accounted for by lower efficiency values for protons acceleration and $\gamma$-ray production.

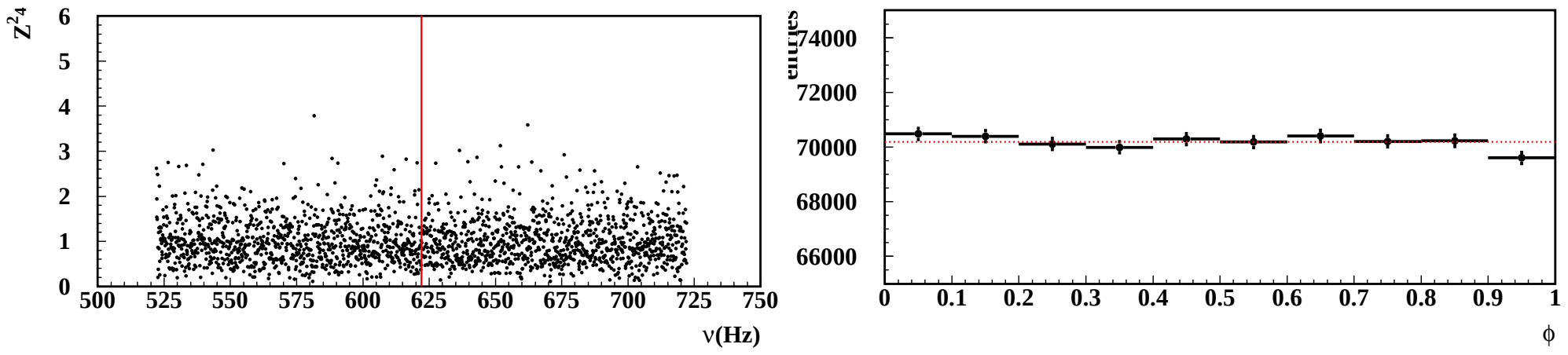

**Figure 1.** On the left, $Z_4^2$ test for a wide range of frequencies around the expected frequency of the millisecond pulsar, marked with a red line. The size of the bin is defined by 1 IFS to make sure that the folding is done for independent frequencies. On the right, the light curve at the expected frequency is represented. The result is compatible within the errors to a flat distribution (red dots line)

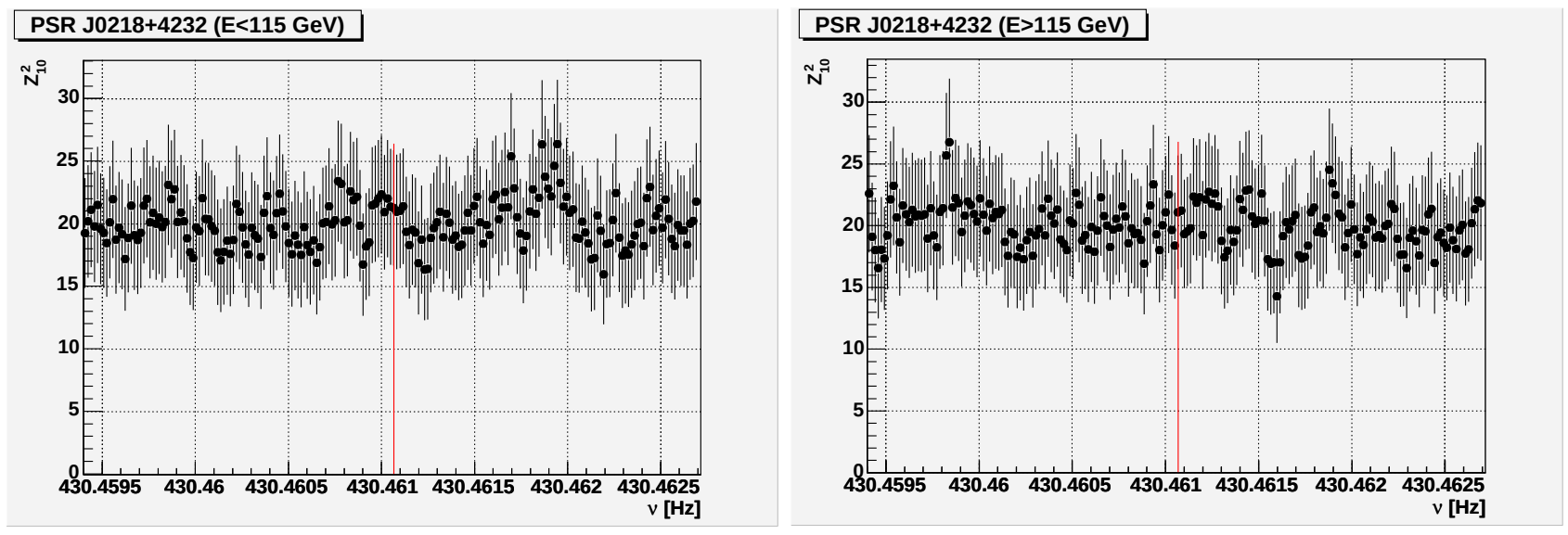

**Figure 2.** On the left, $Z_{10}^2$ test for a wide range of frequencies around the expected frequency of the millisecond pulsar, marked with a red line, for events below $\sim$115 GeV . On the right, the same test for events at high energies.

# Observation of $\gamma$-ray emission above 200 GeV from the AGN 1ES1959+650 during low x-ray and optical activity

N. Tonello[a], D. Kranich[b], D. Paneque[a] and R.M. Wagner[a]
on behalf of the MAGIC Collaboration
*(a) Max-Planck-Institut für Physik, Föhringer Ring, 6, D-80805 München, Germany*
*(b) Department of Physics, University of California, One Shields Avenue, Davis, CA 95616-8677*
Presenter: D.Paneque (dpaneque@mppmu.mpg.de), ger-paneque-D-abs1-og23-oral

The AGN 1ES1959+650 has been observed with the MAGIC telescope in September and October 2004, at the end of the commissioning phase. The MAGIC Telescope is currently the IACT with the lowest energy threshold, which is well suited for observations of weak $\gamma$-ray sources with soft spectra, like AGNs in a quiescent state. During the first observations by MAGIC, 1ES1959+650 was in a low state of activity both in optical and in X-ray wavelengths. The preliminary analysis of VHE $\gamma$-ray data showed a highly significant $\gamma$-ray signal, while no significant variation of emission during the examined period has been found. The results of the analysis and the comparison with previous observations will be presented.

## 1. Introduction

The first $\gamma$-ray signal at very high energy coming from the Active Galactic Nucleus 1ES1959+650 was reported in 1998 by the Seven Telescope Array in Utah, with a 3.9 $\sigma$ significance [19]. Later, other successful observations of the source were reported, but the observed flux was weak in both, $\gamma$-rays and in X-rays.

The AGN 1ES1959+650 is part of an elliptic galaxy at a redshift distance z = 0.047. Observing the source from 2000 until early 2002, the HEGRA collaboration reported only a marginal signal [14]. In May 2002, the X-ray emissions of the source had increased. Both the Whipple [13] and HEGRA [1] collaborations subsequently confirmed also a higher VHE $\gamma$ emission. Further periods of high $\gamma$-ray activity followed in the same year. An interesting aspect of the flaring activities in 2002 was the observation of an orphan flare recorded in June with the Whipple telescope [16, 9], and confirmed by HEGRA [20]: high $\gamma$-ray activity in the absence of high activity in X-rays. Non observation of a significant X-ray activity could be interpreted by the suppression of electron acceleration and inverse Compton scattering as production mechanism for VHE $\gamma$-rays in favor of hadronic models, and would thus deserve special interest. The models for the hadronic component of the 1ES1959 jet (see for example [4]) lead us to also check for neutrino emissions coming from this source. The AMANDA collaboration, operating a neutrino telescope in the southern hemisphere, reported at the recent Cherenkov 2005 meeting at Palaiseau [2] the observation, over a total observation period of four years, five neutrino events, of which three were clustered over a short period coinciding with 1ES1959 flares and one is coincident with the orphan flare. While this observation is statistically insignificant and no more than an indication, even absence of neutrino detection would not rule out the hadronic model in explaining the $\gamma$ emission. The predicted neutrino flux simply is below the sensitivity of present-day neutrino detectors [11]. Future neutrino detectors and extended collaboration between experiments should be able to shed more light on this question.

The MAGIC telescope represents a new generation of IACTs for $\gamma$-ray astronomy. Its design has been optimized to achieve a low energy trigger threshold (eventually down to 30 GeV at zenith), never reached with previous IACTs. The low threshold will make it the instrument of choice for the study of faint VHE $\gamma$-ray sources such as pulsars, medium redshift AGNs, etc. The MAGIC mirror has a diameter and focal length of 17m, its camera is made of 576 hemispherical photo-multiplier tubes with specially shaped light collectors and coated with a diffuse lacquer that enhances their quantum efficiency [6, 7]. The camera has a field of view (FOV) of 3.5°. For more details on the MAGIC telescope and its performance see [8]. The MAGIC telescope is located in the Canary Island of La Palma (28.2°N, 17.8°W, at 2225 m asl). From this location, 1ES1959 is visible from May to October under a zenith angle of 36° at culmination.

## 2. Analysis of the recorded data

This analysis is restricted to $\gamma$s with an energy threshold of $\sim$180 GeV, estimated from the Monte Carlo (MC) $\gamma$ energy peak after analysis. At such energies, we can follow classical techniques for discriminating hadronic and electromagnetic showers, which have been pioneered by Whipple, and described in [10]. The camera image is parameterized to obtain several test statistics describing the image shape and orientation (image parameters or discriminant quantities [12]) in the camera. The number of parameters used in this analysis is eight. The image parameters are used to reject hadronic background events by means of cuts that discriminate between $\gamma$- and hadron-induced images. The parameters also permit reconstructing the arrival direction and the energy of the original $\gamma$-ray.

The period in 2004 where the 1ES1959 data has been taken corresponds to the end of the MAGIC commissioning phase. Generally, the Crab Nebula with its very stable flux is considered a reference source (standard candle) for VHE $\gamma$-ray astronomy. For that reason, Crab data observed with MAGIC were selected such as to match telescope operation conditions, in time and zenith angle range (36° - 47°), to those during the observation of 1ES1959. So called off-source data are collected by pointing the telescope to a sky section near the source, where no $\gamma$-ray signal is expected in the field of view. These data are used as a crosscheck of the recorded cosmic ray background. After quality cuts, $\sim$6.5 hours of observation data (before dead time subtraction) from 1ES1959 remained ($\sim$4.1 M events) and were analyzed. The optimal subspace of image parameters for the $\gamma$-hadron separation was obtained with the Random Forest method [3], using MC $\gamma$ and hadrons recorded during normal ON source data taking. The cut in the discriminant variable provided by the Random Forest (the so-called *hadronness*) was optimized to obtain a signal with the maximum significance from the $\sim$2 hours of Crab Nebula data observed at the same zenith angle. These optimized cuts were then applied to the entire 1ES1959 data sample, without further optimization. Two of the image parameters are of general interest: the variable SIZE, given in number of photons of wavelength between 290 nm and 650 nm entering the camera, is to first order proportional to the energy of the incoming $\gamma$, and ALPHA, the angle between the image major axis and the line connecting the center of gravity of the image with the source position in the camera, shows most clearly the existence of a signal. ALPHA is not used for the optimization process; instead, we derive from the resulting ALPHA distribution the significance of the signal (using [17], formula 17). Finally, data with ALPHA $>$ 9° have been rejected.

## 3. Results

### 3.1 Alpha plot and comparison with Crab Nebula

In figure 1 (left) we show the distribution of the image parameter ALPHA for the Crab Nebula together with the normalized distribution of the off-source data, after a selection of events with SIZE $>$ 1800 photons ($\sim$ 325 photo-electrons), for an energy threshold $\sim$ 300 GeV. At lower SIZE cuts (required to reach the energy threshold of 180 GeV) the alpha distribution is wider, due to a worse reconstruction of the image orientation and larger fluctuations in the atmospheric shower development.

In the right diagram, the ALPHA distribution of the 1ES1959 data sample is shown, after the same parameter cuts. The significance of the 1ES1959 detection is $\sim 8\sigma$, with 252 excess events in $\sim$6 hours of effective observation time, while the signal from Crab Nebula corresponds to $\sim$24 sigma and 583 excess events in $\sim$2 hours for the same analysis. A coarse estimate yields a flux of 0.14 crab units.

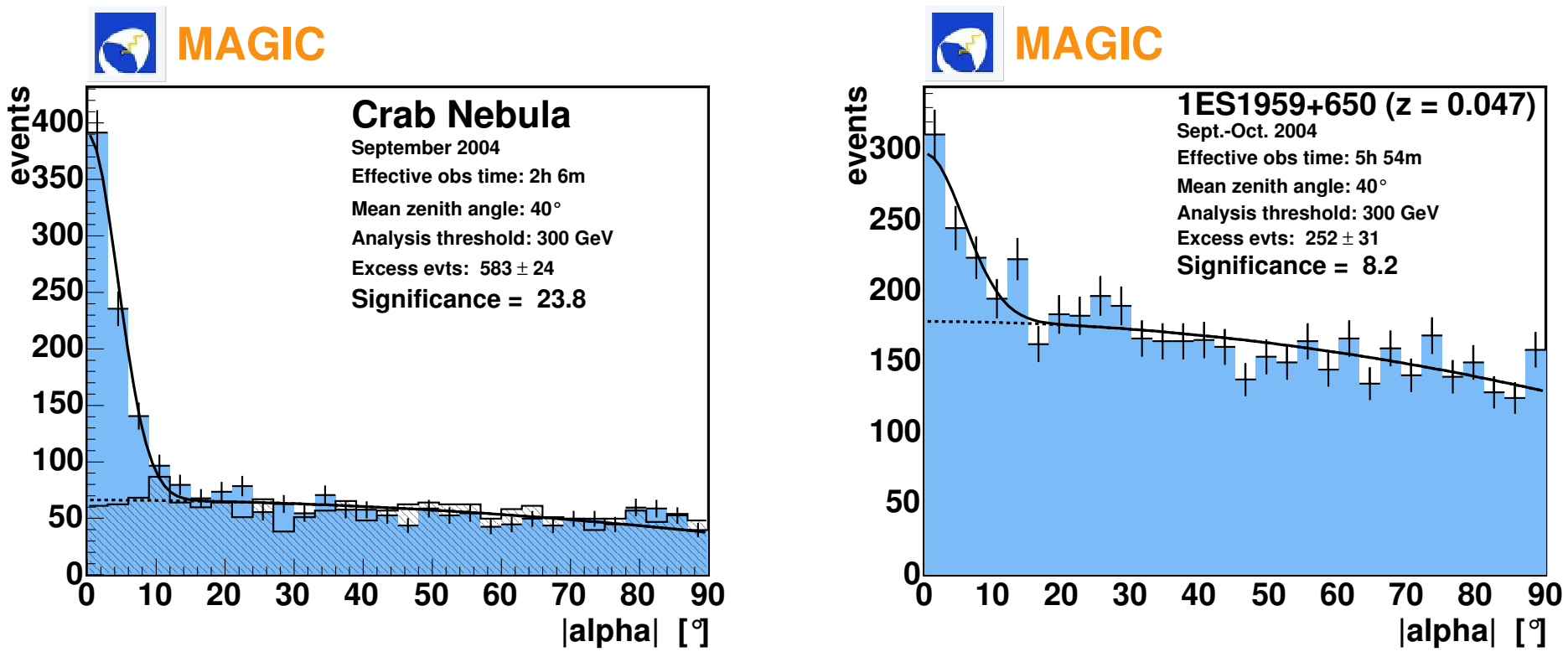

**Figure 1.** ALPHA plots after $\gamma$-hadron separation. Left: Crab, Off data histogram is superimposed for comparison. Right: 1ES1959. The background has been determined with a fit in the region of ALPHA $> 20^\circ$, with a polynomial function of second order.

VHE $\gamma$ ray detection of blazars mostly refer to blazars during a flaring state in other wavelength bands, in particular X-rays [18, 5]. With 6 hours observation time, only modest tests of the flux variation are possible. The data sample of 1ES1959 has been subdivided into 3 sub-samples of $\sim$2 hours each, one sample in September and two in October. The results did not show any significant flux differences during these 3 periods, indicating that the source was basically in the same state during the time covered by our observation.

A strong $\gamma$-ray emission from an AGN leads to the question whether the source was active also at other wavelengths. In case of one zone Synchrotron Self Compton models (see for example [15]), one should also observe a significant activity in the X-ray domain. Indeed, time correlations of the $\gamma$-ray and X-ray fluxes are normally seen in other $\gamma$-ray emitting AGNs. Information of X-ray and optical activity level are based on published RXTE-ASM X-ray flux data (http : //heasarc.gsfc.nasa.gov/xte weather/), the optical light curve being provided by the Tuorla Observatory Blazar Monitoring Program (http : //users.utu.fi/kani/1m/1ES 1959 + 650.html). No strong activity in X-ray or in optical was observed during the period of our VHE $\gamma$-ray detection.

### 3.2 Preliminary 1ES1959 VHE spectrum and comparison with the Crab spectrum

For the spectral flux determination, we used a simple approach of comparing the spectrum of 1ES1959 to that of the Crab nebula, which has been measured by many experiments from data above 400 to 500 GeV, and with MAGIC down to 100 GeV [21].

There is strong evidence of the spectrum of the AGN being steeper than that of the Crab in the same energy range. 1ES1959 is at $\sim$20% of the Crab level at about 200 GeV, while at higher energies ($\sim$ 2 TeV ) the flux drops to $\sim$6% Crab. This observation is compatible with the results of the HEGRA System measurement in a higher energy range of the $\gamma$-ray emissions of 1ES1959 in its steady state of (5.3 $\pm$ 1.1) % Crab [1]. However, our results are still preliminary; a full spectral analysis will be presented elsewhere.

## 4. Conclusions

The AGN 1ES1959 has been clearly detected with the MAGIC telescope after a few hours of observation in September - October 2004, at a mean zenith angle of 40°. During that period, the AGN was in quiescent state both in X-rays and at optical wavelengths.

For an analysis threshold of 300 GeV, the preliminary results yield a 1ES1959 signal at a significance level of $8\sigma$. A simple flux comparison with Crab Nebula in different energy bins from 180 GeV up to 2 TeV indicates that 1ES1959 drops from ∼20% around 200 GeV to ∼6% Crab flux at around 2 TeV. The full spectral analysis is still in progress, and will be presented elsewhere. The present observation is characterized by lack of strong time variability in $\gamma$-rays, as well as in the X-rays and the optical domain. Therefore, we assumed that 1ES1959 was observed in quiescent state. It is worth noticing that this is the first time ever that this AGN is significantly detected in the VHE domain in quiescent state in only $\sim 6$ hours of observation time. This experimental result can be seen as a confirmation that MAGIC is suited for observing weak $\gamma$-ray sources with soft spectra, like AGNs in non-flaring state.

## 5. Acknowledgements

We would like to thank the IAC for the excellent working conditions. The support of the German BMBF and MPG, the Italian INFN and the Spanish CICYT is gratefully acknowledged.

# Observations of the Crab nebula with the MAGIC telescope

R. M. Wagner[a], M. Lopez[b], K. Mase[a,f], E. Domingo–Santamaria[c], F. Goebel[a], J. Flix[c], P. Majumdar[a], D.Mazin[b], A. Moralejo[d], D. Paneque[a], J. Rico[c], and T. Schweizer[e]
on behalf of the MAGIC collaboration

*(a) Max-Planck-Institut für Physik, Föhringer Ring 6, D-80805 München, Germany*
*(b) Universidad Complutense de Madrid, Facultad de Ciencias Fisicas, E-28040 Madrid, Spain*
*(c) Institut de Fisica d'Altes Energies, E-08193 Bellaterra, Spain*
*(d) Dipartimento di Fisica Galileo Galilei and INFN Padova, Via Marzolo 8, I-35131 Padova, Italy*
*(e) Humboldt-Universität zu Berlin, Institut für Physik, Newtonstrasse 15, D-12489 Berlin, Germany*
*(f) now at Chiba University, Department of Physics, Yayoi-tyo 1-33, Inage-ku, Chiba-shi, Chiba-ken, Japan 263-8522*

Presenter: R. M. Wagner (robert.wagner@mppmu.mpg.de), ger-wagner-R-abs1-og22-oral

During and shortly after the telescope commissioning the MAGIC collaboration observed the Crab nebula. Its steady flux of gamma rays provides good means for studying the telescope performance. Here we present results obtained from these observations. Emphasis is put on the stability of the flux determination during periods with different telescope performances and on describing new analysis methods used to extract signals in the low energy region. The analysis is restricted to energies above 100 GeV, since details in the $\gamma$/hadron separation in the low energy region and the telescope performance require more studies.

## 1. Introduction

The 17 m diameter MAGIC telescope[1, 2], located on the Canary island of La Palma (28.8° North, 17.8° West, 2200 m a.s.l.), is a new generation, high performance air Cerenkov telescope for very high energy (VHE) $\gamma$–astronomy. The main design goal was to achieve a very low ($\sim$30 GeV) trigger threshold in order to bridge the energy gap between the satellite borne $\gamma$–detectors and ground-based Cerenkov telescopes of the last decade. MAGIC was commissioned in 2004 and started first observations in the same year. Emphasis was put on the verification of established sources in order to study the telescope performance in detail [3]. As the MAGIC telescope construction is based on many novel elements, untried up to now, the current analysis is restricted to the energy range above approx. 100 GeV, e.g. to shower images of at least 150–200 photoelectrons. MAGIC has already detected four well-established $\gamma$-sources, namely the Crab nebula, the AGN Mkn 421 [4], Mkn 501 [1], and 1ES1959+650 [5], and has seen evidence for a few more sources [1].

## 2. The Crab Nebula

The Crab nebula is the remnant of a supernova explosion that occurred in 1054. In 1989, VHE $\gamma$ emission was reported by the Whipple collaboration [6]. It was the first source detected at TeV energies employing the IACT technique and it exhibits a stable and strong $\gamma$-emission. It therefore is frequently used as the "standard candle" in VHE $\gamma$-astronomy. The Crab nebula has been observed extensively in the past over a wide range of wavelengths, covering the radio, optical and X-ray bands, as well as high-energy regions up to nearly 100 TeV [7]. Nevertheless, quite some new physics results are expected in the VHE domain, namely the spectrum showing an Inverse Compton (IC) peak close to 100 GeV, a cut-off of the pulsed $\gamma$-emission somewhere between 10 and 100 GeV, and the verification of true flux stability down to the percent level. Currently the VHE $\gamma$-emission is very well described by electron acceleration followed by the IC scattering of photons generated by synchrotron radiation (SSC model [8]). Probing the presence/absence of a small contribution of VHE $\gamma$s produced in hadronic interactions is a challenge for experimenters.

## 3. Data Analysis

**Data Sample.** The data analyzed here were taken during September and October 2004 and in January 2005. A total of 2.8M events in 2004 and 4.5M events from the 2005 observations were used. The analysis is restricted to a sample of low zenith angle observations (ZA$< 30^\circ$). Quality checks were performed in order to reject runs with unstable trigger rates due to variable atmospheric conditions. The overall observation time of the sample analyzed corresponds to 13 hours on-source.

**Calibration, Flatfielding and Event Reconstruction.** The camera has been flatfielded and the gains of the PMTs calibrated on a run-by-run basis using a fast UV LED pulser [9]. For obtaining the conversion factor from ADC counts to photoelectrons the excess noise factor method [10] was used. The telescope QE was calculated from the optical parameters of the different components and verified from Muon ring data [11]. After calibration, a cleaning algorithm was applied to the shower images to remove the contribution of the night sky light background, using a cut in the number of photoelectrons per PMT pixel. Finally, the images were parameterized in terms of the well-known Hillas parameters [12].

**Rejection of the Hadronic Background.** For $\gamma$/hadron separation we used a technique based on the Random Forest (RF) method [13, 14]. The conceptual difference compared to dynamical or scaled cuts in image parameters is that instead of applying an independent cut to each image parameter, the RF method uses all the parameters simultaneously, taking into account interdependencies and scaling of the image parameters, e.g. with energy, impact parameter, and zenith angle automatically. Monte Carlo $\gamma$s and real hadronic background data have been used as training samples. The Hillas parameters *Size*, *Dist*, *Width*, *Length*, *Conc* and *Asym* have been used in the training. The RF method tags each event with the so-called hadronness ($h$), which is related to the event's probability to be of hadronic origin. An appropriate cut in $h$ yields a sample retaining most of the $\gamma$-candidates while suppressing a large fraction of the hadronic background. Finally, a cut in *Alpha*, the angle between the shower major axis and the line connecting the shower COG with the source location in the FOV, allows to further suppress background (overall cut efficiency: $\approx 60\%$). At lower energies the discrimination power between $\gamma$s and hadrons degrades, because $\gamma$ and hadron images look more and more similar. In addition, the *Alpha* distributions broaden, since the shower images contain fewer PMT pixels and thus the reconstruction of the shower direction degrades. The cuts in $h$ and *Alpha* are therefore chosen as a function of energy. Below 100 GeV $\gamma$-background separation becomes more difficult because the Earth's magnetic field gradually distorts the shower images, the fluctuations of the shower structure increase, more compact and smaller images occur and contributions from cosmic electrons and $\pi^0$ induced showers from hadronic interactions increase where all other secondary particles are below the Cerenkov threshold. All these effects require many more studies for a good $\gamma$-background rejection below 100 GeV. It should also be mentioned that the majority of muon-induced images could easily be rejected, but quantitative numbers are not yet available.

**Alpha distribution and Sky Map.** In Fig. 1 we show two *Alpha* distributions for the data subsample of October 10, 2004, one above a *Size* of 360 photoelectrons (ph.e.), yielding a signal of $20.4\sigma/\sqrt{h}$, and a second distribution above a Size of 2000 ph.e. Note that the latter $\gamma$-sample is nearly background free. Fig. 2 shows an excess sky map of the corresponding sky region obtained with the *Disp*–Method [15].

**Energy estimation.** The recorded light content of $\gamma$-showers with an impact parameter $\lesssim$ 120m (*Dist* $< 1^\circ$) is in first order proportional to the initial energy. In order to estimate the energy of each event, we trained RFs for each energy bin using log(*Size*), *Dist*, *Width*, *Length*, log(*Size*/(*Length* x *Width*)), *Conc*, *Leakage*, and the zenith angle. After training, for each event and each energy bin a quantity analogous to the hadronness is computed, yielding the probability of this event being in the respective bin. By weighting all possible energies with these probabilities one gets a continuous energy estimation (cf. fig. 3).

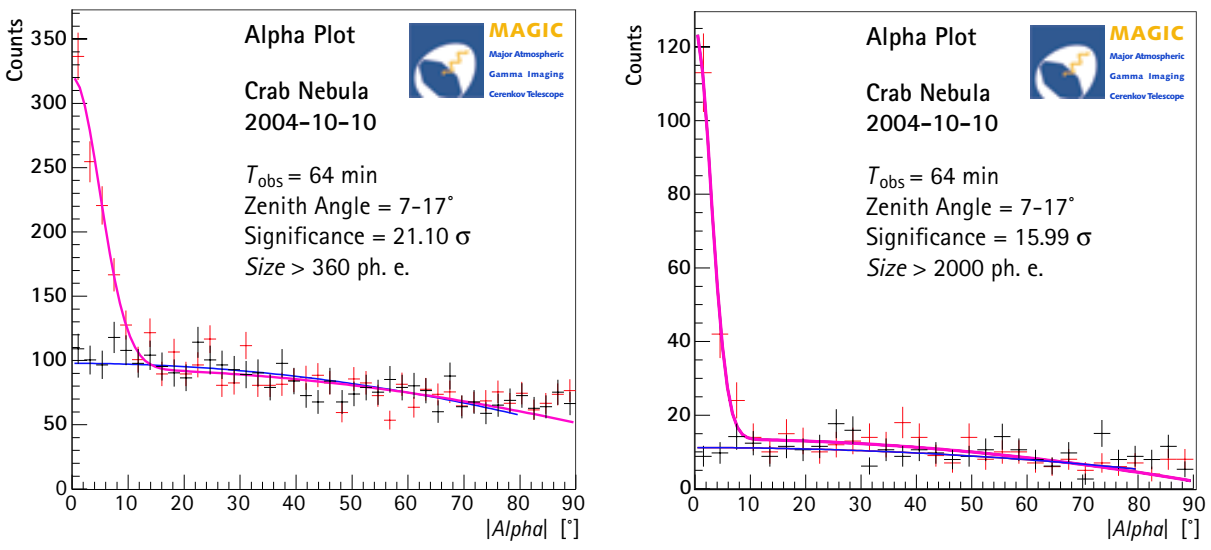

**Figure 1.** *Alpha* distribution for 64 minutes of observation time.

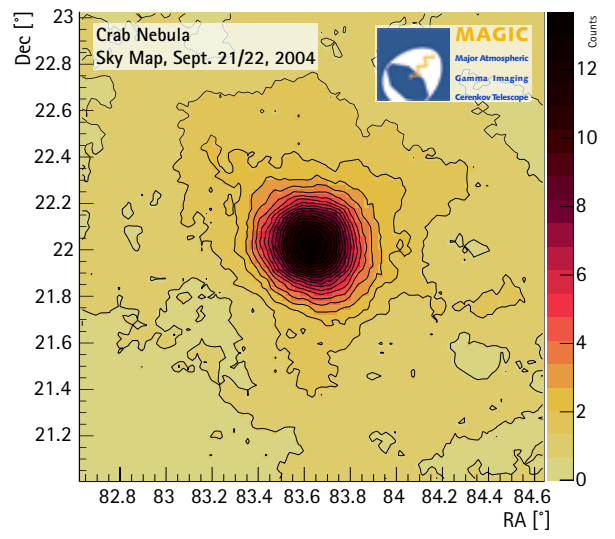

**Figure 2.** *Disp*–Sky map for the Crab nebula.

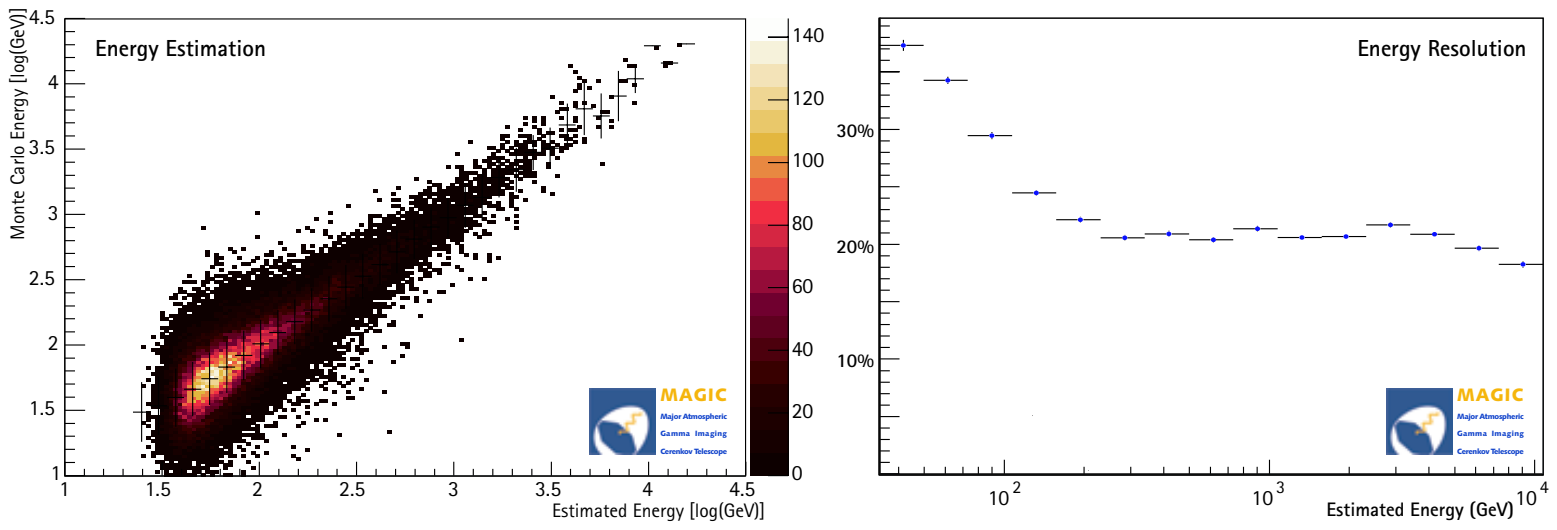

**Figure 3.** Left: Estimated energy vs. true energy, for Monte Carlo events. Right: Energy resolution as a function of (estimated) energy.

**Flux stability and Spectrum.** For calculating the true energies from the estimated energies, we apply a robust "spillover correction": We compute the ratio $c_i = N_\gamma^{(E_{\text{true}} \in B_i)} / N_\gamma^{(E_{\text{estimated}} \in B_i)}$ for each bin $B_i$ in estimated energy. When multiplied with the number of excess events $N_i$, it yields the number $N_i^{\text{true}}$ of events belonging to the respective bin in true energy. For the final flux calculation the data have to be corrected for various losses, such as for different detector inefficiencies, dead–time effects, atmospheric transmission corrections etc. Still, not all corrections could be taken into account. We estimate the systematic uncertainty in the flux to be $\approx$35%.

Fig. 4 shows the integral flux of the Crab nebula above 200 GeV for the individual days of this analysis. Fig. 5 shows the differential spectrum of the Crab nebula. A power–law fit between 300 and 3000 GeV yielded a spectral index of $2.58 \pm 0.16$. In agreement with expectance, the measured data points below 300 GeV lie below the extrapolated power law.

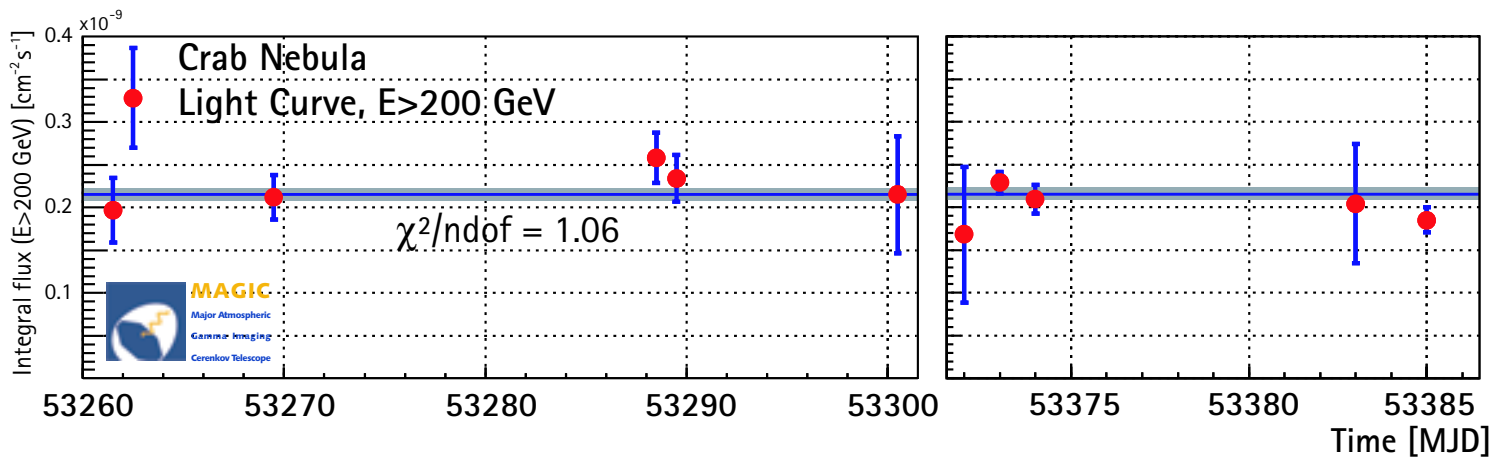

**Figure 4.** Integral flux of the Crab nebula for all days considered in this analysis. A fit assuming a constant flux gave a $\chi^2/\text{ndof} = 1.06$.

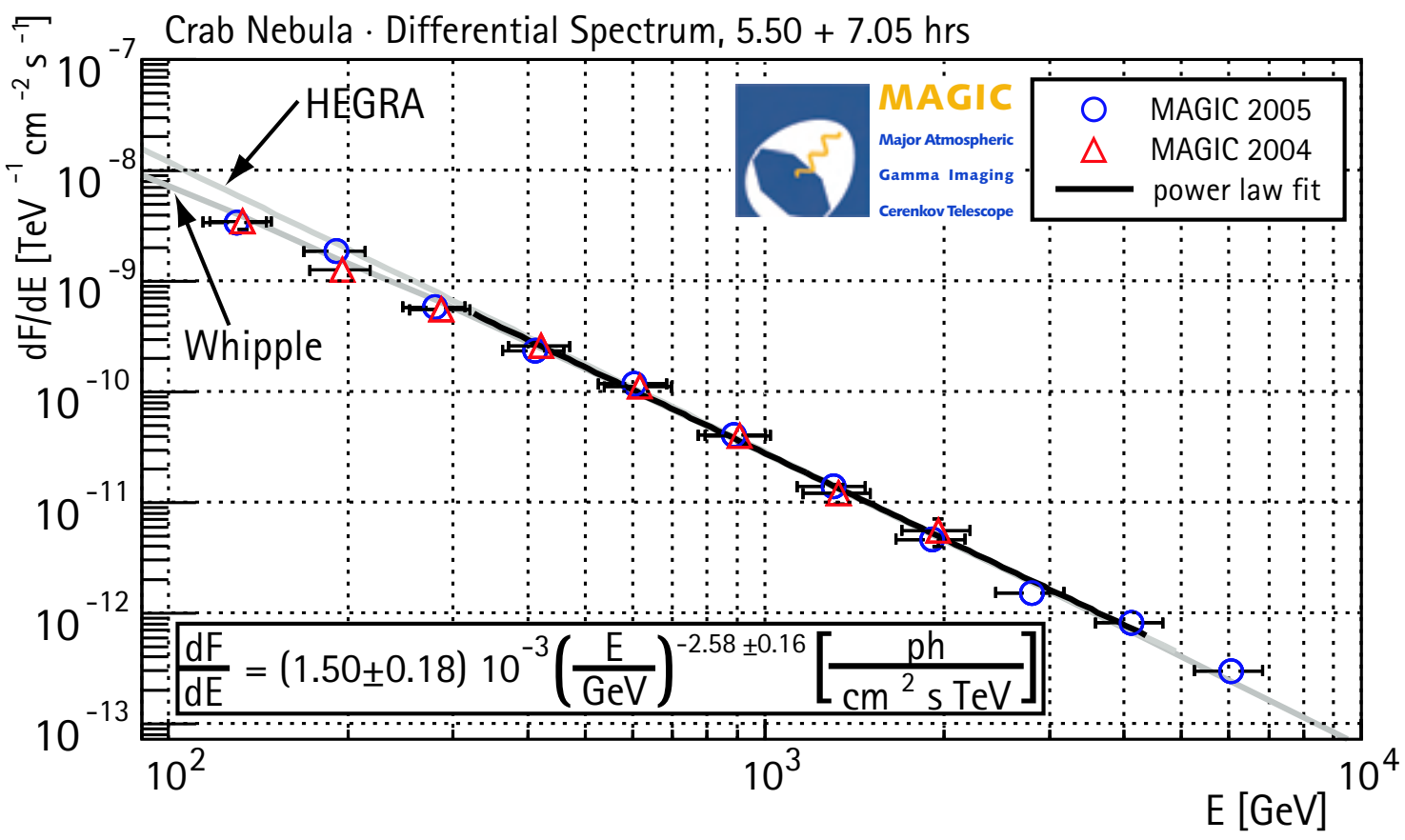

**Figure 5.** Observed differential crab spectra for the 2004 and 2005 datasets. We show a power–law fit to the combined data between 300 GeV and 2 TeV as well as the fit to HEGRA data [7] and the parametrization of Whipple data [12].

## 4. Conclusions

The MAGIC telescope has been taking data since mid 2004. As a first target of observation the Crab nebula was observed to test the gross performance of the telescope. An analysis between 100 and 4000 GeV confirmed expectations and gave good agreement with observations of other groups. We determined an energy resolution of $\approx 20\%$ at energies in the above quoted energy range. Although run–in data were used for this analysis, we observed good flux stability. The next steps in analysis will focus on the region below 100 GeV and on the reduction of the many systematic errors.

## Acknowledgments

We would like to thank the Instituto de Astrofísica de Canarias for excellent working conditions. The support of the German BMBF and MPG, the Italian INFN and the Spanish CICYT is gratefully acknowledged.